\documentclass[showpacs,aps,prd,reprint,superscriptaddress,,nofootinbib,longbibliography]{revtex4-2}
\usepackage[colorlinks=true, pdfstartview=FitV,
bookmarks=true, bookmarksnumbered=true, breaklinks]{hyperref}
\usepackage[dvipdfmx]{graphicx}
\usepackage{amsmath,amssymb,bm,color,longtable,mathrsfs,slashed,comment,tikz,dcolumn,braket}

\definecolor{darkviolet}{rgb}{0.58, 0.0, 0.83}
\definecolor{electricultramarine}{rgb}{0.25, 0.0, 1.0}
\definecolor{brightpink}{rgb}{1.0, 0.0, 0.5}
\hypersetup{linkcolor=brightpink, citecolor=darkviolet, urlcolor=electricultramarine}

\definecolor{lime}{HTML}{A6CE39}
\DeclareRobustCommand{\orcidicon}{
	\hspace{-3mm}
	\begin{tikzpicture}
	\draw[lime, fill=lime] (0,0) 
	circle [radius=0.16] 
	node[white] {{\fontfamily{qag}\selectfont \tiny ID}};
	\draw[white, fill=white] (-0.0625,0.095) 
	circle [radius=0.007];
	\end{tikzpicture}
	\hspace{-3mm}
}

\foreach \x in {A, ..., Z}{\expandafter\xdef\csname orcid\x\endcsname{\noexpand\href{https://orcid.org/\csname orcidauthor\x\endcsname}
			{\noexpand\orcidicon}}
}

\begin{document}

\title{Dual chiral density wave induced oscillating Casimir effect}

\author{Daisuke Fujii\orcidA{}}
\email[]{daisuke@rcnp.osaka-u.ac.jp}
\affiliation{Advanced Science Research Center, Japan Atomic Energy Agency (JAEA), Tokai, 319-1195, Japan}
\affiliation{Research Center for Nuclear Physics, Osaka University, Ibaraki 567-0048, Japan}

\author{Katsumasa~Nakayama\orcidB{}}
\email[]{katsumasa.nakayama@riken.jp}
\affiliation{RIKEN Center for Computational Science, Kobe, 650-0047, Japan}

\author{Kei~Suzuki\orcidC{}}
\email[]{k.suzuki.2010@th.phys.titech.ac.jp}
\affiliation{Advanced Science Research Center, Japan Atomic Energy Agency (JAEA), Tokai, 319-1195, Japan}


\begin{abstract}
The Casimir effect is known to be induced from photon fields confined by a small volume, and also its fermionic counterpart has been predicted in a wide range of quantum systems.
Here, we investigate what types of Casimir effects can occur from quark fields in dense and thin quark matter.
In particular, in the dual chiral density wave, which is a possible ground state of dense quark matter, we find that the Casimir energy oscillates as a function of the thickness of matter.
This oscillating Casimir effect is regarded as an analog of that in Weyl semimetals and is attributed to the Weyl points in the momentum space of quark fields.
In addition, we show that an oscillation is also induced from the quark Fermi sea, and the total Casimir energy is composed of multiple oscillations.
\end{abstract}

\maketitle


\section{\label{introduction}Introduction}

The Casimir effect, proposed by Casimir~\cite{Casimir:1948dh}, is crucially important for understanding small-volume physics in quantum field theory (see Refs.~\cite{Plunien:1986ca,Mostepanenko:1988bs,Bordag:2001qi,Milton:2001yy,Klimchitskaya:2009cw} for reviews).
Casimir predicted that the decrease of the zero-point energy of photon fields by two parallel plates would cause an attractive force for the plates, which are the so-called Casimir energy and the Casimir force.
The Casimir force was experimentally verified about fifty years later~\cite{Lamoreaux:1996wh,Bressi:2002fr}.

Beyond academic interest, the engineering application of the Casimir effect to nanotechnology (Casimir engineering) has recently attracted much attention~\cite{Gong:2020ttb}.
The most typical feature of the Casimir effect is an attractive force.
On the other hand, when one tunes the permittivity and/or permeability of plates and medium, a repulsive force can be also realized~\cite{Dzyaloshinskii:1961sfr,Boyer:1974buu,Kenneth:2002ij,Munday:2009fgb}.
In contrast to such attractive or repulsive Casimir effects, the third type of Casimir effect would be interesting, where ``third" means that attraction or repulsion is not fixed.
For example, under a setup, the sign of Casimir energy flips from attraction to repulsion as the separation distance increases, which may be called the {\it sign-flipping Casimir effect} (e.g., see Refs.~\cite{Henkel:2005,Rosa:2008zza,Grushin:2010qoi,Tse:2012pb,Wilson:2015,Fukushima:2019sjn}).
As another example, the value of Casimir energy can oscillate as a function of distance, which may be called the {\it oscillating Casimir effect} (e.g., see Refs.~\cite{Fuchs:2007,Wachter:2007,Kolomeisky:2008,Zhabinskaya:2009,Chen:2011xtv,Chen:2011JStatPhys,Han:2015,Jiang:2018ivv,Ishikawa:2020ezm,Ishikawa:2020icy,Mandlecha:2022cll,Nakayama:2022fvh,Nakata:2023keh}).
Among them, Ref.~\cite{Nakayama:2022fvh} found that it occurs inside Weyl semimetals, where the origins of the oscillation are Weyl points (WPs) at finite momenta in the dispersion relations of Weyl fermions.
Such new types of Casimir effects are not only of theoretical interest but also will be important for Casimir engineering.

The Casimir effect is also important for elucidating quark and gluon dynamics described by quantum chromodynamics (QCD) in a small volume.
For example, (i) as an {\it ab initio} method for solving QCD, numerical simulations of lattice QCD are done in finite volume (e.g., in a box of a few fm size), and finite-volume effects must be understood.
Since the finite-volume effect for zero-point energy is nothing but the Casimir effect, its understanding is helpful for interpretations of results in small-volume simulations.
(ii) In relativistic heavy-ion collision experiments, quark-gluon plasma is produced as a fireball with a size of a few fm. 
Therefore, we must well understand the contribution of the Casimir effect to physics inside the fireball and near its boundary.
(iii) In the interiors of neutron stars, dense quark matter as well as nuclear matter may exist.
Depending on the microscopic density profile inside stars, there may be small regions of quark matter.

\begin{figure}[b!]
    \centering
    \begin{minipage}[t]{1.0\columnwidth}
    \includegraphics[clip,width=1.0\columnwidth]{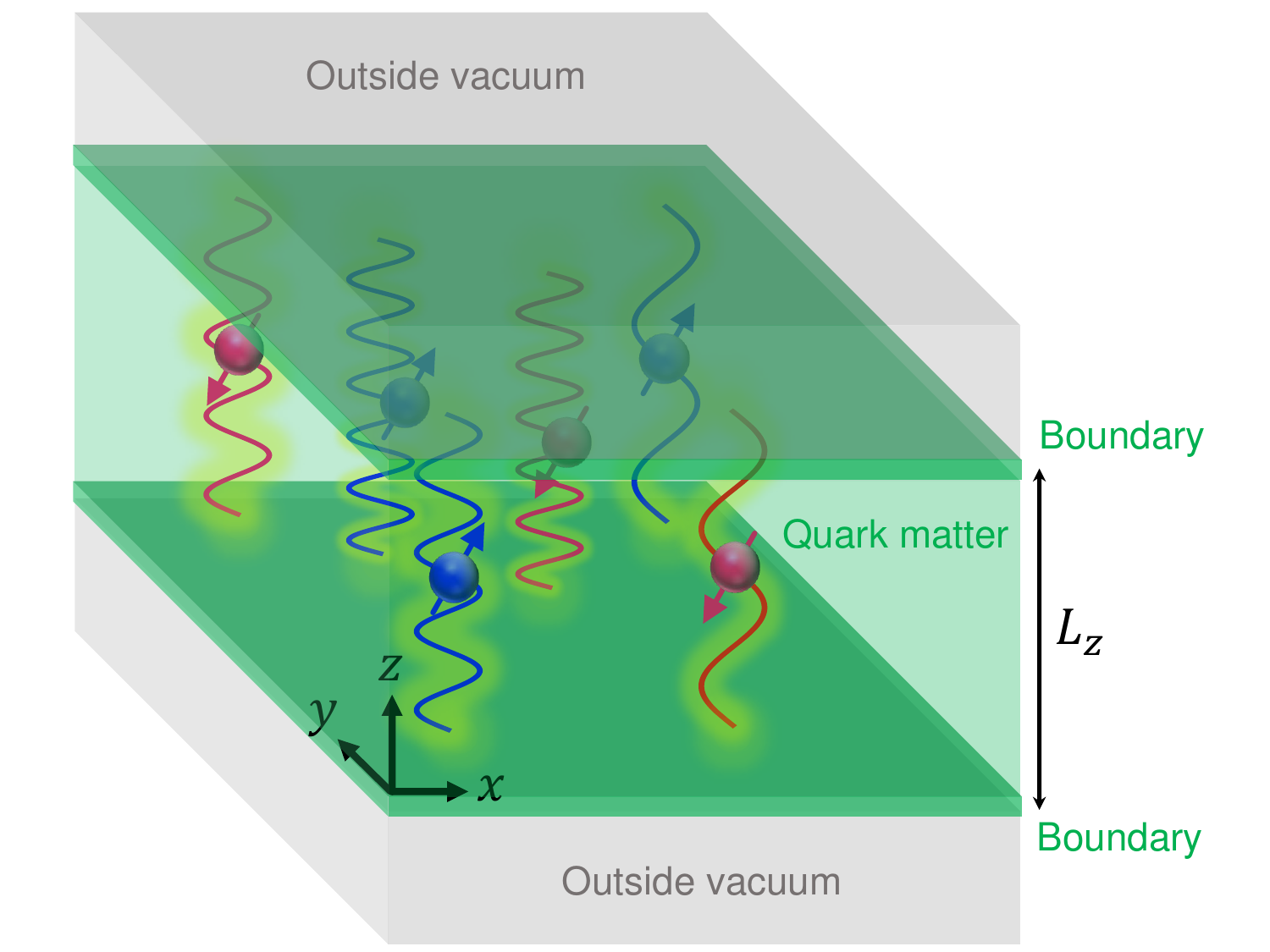}
    \end{minipage}
\caption{
Schematics of the Casimir effect for quark fields inside thin quark matter, where the quark matter is confined between boundary conditions with a thickness $L_z$.
The ``outside vacuum" should be regarded as an arbitrary vacuum consistent with the boundary conditions.
}
\label{fig:schematic}
\end{figure}

Under these motivations, this paper focuses on what types of Casimir effects can occur in various phases of quark matter where the thickness of $z$ direction is extremely short (i.e., ``thin" quark matter) as illustrated in Fig.~\ref{fig:schematic}.
In particular, in this paper, we propose a QCD counterpart of the oscillating Casimir effect predicted in Weyl semimetals.
This is realized for fermion fields in the dual chiral density wave (DCDW) phase~\cite{Dautry:1979bk} (also see Refs.~\cite{Broniowski:1989tw,Kutschera:1989yz,Broniowski:1989fz,Broniowski:1990gb,Kutschera:1990xk,Kato:1992iw,Kotlorz:1994sk,Sadzikowski:2000ap,Takahashi:2001hp,Takahashi:2001jq,Sadzikowski:2002iy,Takahashi:2002md,Tatsumi:2004dx,Nakano:2004cd,Sadzikowski:2006jq,Bringoltz:2006pz,Nickel:2009wj,Maedan:2009yi,Carignano:2010ac,Frolov:2010wn,Partyka:2010em,Abuki:2011pf,Carignano:2012sx,Karasawa:2013zsa,Muller:2013tya,Heinz:2013hza,Moreira:2013ura,Tatsumi:2014cea,Carignano:2014jla,Tatsumi:2014wka,Hayata:2014eha,Harada:2015lma,Lee:2015bva,Nishiyama:2015fba,Carignano:2015kda,Carlomagno:2015nsa,Yoshiike:2015tha,Heinz:2015lua,Suenaga:2015daa,Karasawa:2016bok,Adhikari:2017ydi,Yoshiike:2017kbx,Tatsumi:2018ifx,Carignano:2019ivp,Lakaschus:2020caq,TabatabaeeMehr:2023tpt,Pitsinigkos:2023xee} and for related studies and Ref.~\cite{Buballa:2014tba} for a review).
The DCDW phase is a candidate for the ground state of quark/nuclear matter in the density region of $\rho/\rho_0=3\sim6$ with the normal nuclear density $\rho_0=0.16$ ${\rm fm}^{-3}$~\cite{Tatsumi:2004dx,Nakano:2004cd}, and complex dispersion relations of fermions in this phase can be an origin of the oscillating Casimir effect.

In this paper, we aim to investigate types of Casimir effects induced from quark fields in dense quark matter.
To accomplish this, this paper is organized as follows.
In Sec.~\ref{formultaion}, we formulate the zero-point energy and the Casimir energy within the Nambu--Jona-Lasinio (NJL) model~\cite{Nambu:1961tp,Nambu:1961fr} as an effective model of QCD.
In Sec.~\ref{Results}, we show our numerical results and classify the types of Casimir effects in three density regions.
Section~\ref{Conclusions} is devoted to our conclusion and outlook.

\section{\label{formultaion}Formulation}

\subsection{\label{DCDW}Dual chiral density waves}

A chiral density wave is a spatially inhomogeneous chiral condensate, which is represented as a position-dependent Lorentz-scalar condensate made of the Dirac fields, $\psi$ and $\bar{\psi}=\psi^\dagger \gamma_0$: $\langle \bar{\psi}\psi\rangle \propto\cos(\vec{q}\cdot\vec{r})$ with a wave number vector $\vec{q}$ and a position vector $\vec{r}$.
Among various types of chiral density waves, in this study, we focus on the DCDW, where both the scalar and pseudoscalar condensates are position-dependent~\cite{Dautry:1979bk}, 
\begin{eqnarray}
\langle\bar{\psi}\psi\rangle=\Delta\cos(\vec{q}\cdot\vec{r}), \ \ \ \langle\bar{\psi}i\gamma_5\psi\rangle=\Delta\sin(\vec{q}\cdot\vec{r}),
\end{eqnarray}
where $\Delta$ is the amplitude of the DCDW and is regarded as the radius of the chiral circle, $\langle\bar{\psi}\psi\rangle^2+\langle\bar{\psi}i\gamma_5\psi\rangle^2=\Delta^2$.

\subsection{\label{NJL}NJL model}
To investigate the DCDW phase of quark matter, in this work, we use the NJL model~\cite{Nambu:1961tp,Nambu:1961fr} (see Refs.~\cite{Vogl:1991qt,Klevansky:1992qe,Hatsuda:1994pi,Buballa:2003qv} for reviews), which was done in Refs.~\cite{Sadzikowski:2000ap,Sadzikowski:2002iy,Tatsumi:2004dx,Nakano:2004cd} for early studies.
The Lagrangian density of the two-flavor NJL model is written as
\begin{eqnarray}
\mathcal{L}_{\rm NJL}=\bar{\psi}(i\partial\llap{/}+\mu\gamma_0)\psi+G[(\bar{\psi}\psi)^2+(\bar{\psi}i\gamma_5\vec{\tau}\psi)^2],
\end{eqnarray}
where $\mu$ is the quark chemical potential, $G$ is the coupling constant of the four-point interactions, and $\vec{\tau} =(\tau_1,\tau_2,\tau_3)$ is the Pauli matrix for the isospin.
Here, we apply the mean-field ansatz for the DCDW,
\begin{align}
&\langle \bar{\psi}\psi \rangle=\Delta\cos(\vec{q}\cdot\vec{r}),
&&\langle \bar{\psi}i\gamma_5 \tau_3 \psi\rangle=\Delta\sin(\vec{q}\cdot\vec{r}), \\
&\langle \bar{\psi}i\gamma_5 \tau_1 \psi\rangle=0,
&&\langle \bar{\psi}i\gamma_5 \tau_2 \psi\rangle=0,
\end{align}
where $\Delta$ is the amplitude of the DCDW, and $\vec{q}=(0,0,q)$ is the wave number of the DCDW propagating in the $z$ direction. 
Using this ansatz, the mean-field (MF) Lagrangian density is
\begin{align}
\mathcal{L}_{\rm MF}=&\bar{\psi}\big[i\partial\llap{/}+\mu\gamma_0-M\big(\cos(qz)+i\gamma_5\tau_3\sin(qz)\big)\big]\psi \notag \\ 
&-\frac{M^2}{4G} \label{Lagrangian},
\end{align}
where $M=-2G\Delta$.

Using a local chiral transformation (which may be called a Weinberg transformation), $\psi e^{i \gamma_5 \tau_3 qz/2} \to \psi_{\rm W}$, $\bar{\psi} e^{i \gamma_5 \tau_3 qz/2} \to \bar{\psi}_{\rm W}$, the original quark field $\psi$ is redefined as a new field $\psi_{\rm W}$, and the position dependence of the original Lagrangian (\ref{Lagrangian}) is removed in the redefined quark fields.
Then, from the diagonalization of the inverse quark propagator in momentum space, we obtain the four eigenvalues of (quasi-)quarks, $\omega_\pm$ (the positive-energy modes if $\mu=0$) and $\tilde{\omega}_\pm$ (the negative-energy modes): 
\begin{eqnarray}
&&\omega_\pm = E_s-\mu, \label{eq:E+-} \\
&&\tilde{\omega}_\pm =-E_s -\mu, \label{eq:tildeE+-}
\end{eqnarray}
where
\begin{eqnarray}
&&E_s= \sqrt{k^2+M^2+q^2/4\pm\sqrt{(qk_z)^2+M^2q^2}}, \label{eq:Es}
\end{eqnarray}
and $k^2=k_x^2+k_y^2+k_z^2$.
When $q=0$, the dispersion relations return to those of usual massive quarks with a mass $M$, where the two modes with different spins are degenerate.
When $q \neq 0$, the two modes are split, which is labeled by $s=\pm$.
In this work, we focus on the case of $M<q/2$, which is a typical situation of the DCDW phase.\footnote{The case of $M \geq q/2$ is also interesting, but the WP-induced oscillating Casimir effect does not occur.}

Using these eigenvalues, the thermodynamic potential (per a spatial volume $V=L_xL_yL_z$) at temperature $T=1/\beta$ is written as
\begin{align}
\frac{\Omega(T)}{V} =&-N_fN_c\int\frac{d^3 k}{(2\pi)^3} \sum_{s=\pm}\left[E_s +\frac{1}{\beta}\ln\big(1+e^{-\beta(E_s-\mu)}\big) \right. \notag\\
&\left. 
+\frac{1}{\beta}\ln\big(1+e^{-\beta(E_s+\mu)}\big)\right]+\frac{M^2}{4G}. \label{Eq:Omega}
\end{align}
In this work we fix the number of flavors as $N_f=2$ and the number of colors as $N_c=3$.

By taking the zero-temperature limit $T\rightarrow0$ of the thermodynamic potential~(\ref{Eq:Omega}), we obtain
\begin{align}
\frac{\Omega(T\to 0)}{V} \equiv \frac{E_{0}}{L_z}=&-N_fN_c\int\frac{d^3 k}{(2\pi)^3}\sum_{s=\pm}\left[E_s \right. \notag \\
&\left. +\left(\mu-E_s \right)\theta(\mu-E_s)\right]
+\frac{M^2}{4G}. \label{zeropoint}
\end{align}
Here, the first term is regarded as the contribution from the Dirac sea, i.e., the negative zero-point energy from the fermion fields.
The second term with the step function is the contribution from the Fermi sea of quarks.
The third term is the positive energy from the order parameter $M$.
To provide a Casimir-effect-like picture also for the Fermi sea contribution (as well as the Dirac sea), we have denoted Eq.~(\ref{zeropoint}) as $E_0/L_z$ where $E_0$ is the zero-point energy per unit area.

Note that in the traditional treatment of the NJL model, we determine the values of the order parameters $M$ and $q$ by minimizing Eq.~(\ref{zeropoint}) at a fixed $\mu$.
In a finite volume, the thermodynamic potential in a fixed volume is minimized, and the values of $M$ and $q$ should be determined.
This type of analysis is important for studying the phase diagram in finite volume, but it is not the purpose of this work.
The purpose of this work is to investigate the feasibility of the oscillating Casimir effect in dense quark matter.

\subsection{\label{Lifshitz}Casimir energy with Lifshitz formula}

The Casimir energy is defined as a finite-volume effect for the zero-point energy (\ref{zeropoint}).
In a finite volume, the momentum integral with respect to the three-dimensional momentum of quarks in Eq.~(\ref{zeropoint}) is replaced by a discrete sum.
In this paper, we impose the periodic boundary conditions (PBCs) on quark fields at $z=0$ and $z=L_z$, where the $z$ component of quark momentum is discretized as $k_z\rightarrow2n\pi/L_z$ where $n=0,1,\cdots, \infty$.\footnote{The case with boundary conditions for the $x$ or $y$ direction is straightforward, but these boundaries do not induce the WP-induced oscillating Casimir effect since now the DCDW is along the $z$ axis.}
Note that for the PBC, there is no outside vacuum as in Fig.~\ref{fig:schematic}, but the discretization of momentum (and corresponding eigenvalues) can induce a nonzero Casimir energy.

The zero-point energy (\ref{zeropoint}) as the infinite integral and the corresponding infinite sum is divergent (unless an energy or a momentum cutoff is introduced), but the Casimir energy should be finite after using a mathematical technique.
In this paper, we propose two approaches: (i) the Lifshitz formula and (ii) lattice regularizations.

Using the first approach, at zero temperature $T=0$ and zero chemical potential $\mu=0$, the analytical solution for the Casimir energy (per unit area) from quark fields in the DCDW phase can be written as
\begin{align}
E_{\rm Cas}
&=-4N_f N_c\sum_{s=\pm}\int_0 ^\infty\frac{d\xi}{2\pi}\int\frac{dk_xdk_y}{(2\pi)^2}\ln\big[1-e^{-L_z\tilde{k}_z^{[s]}}\big], \notag \\
\tilde{k}^{[\pm]}_z&=\sqrt{k_\perp^2+M^2+\xi^2-\frac{q^2}{4}\mp iq\sqrt{k_\perp^2+\xi^2}}, \label{eq:ECas_PBC}
\end{align}
where the first integral variable is the imaginary part $\xi$ of imaginary frequency $\omega \equiv i\xi$, and $k_\perp^2 \equiv k_x^2+k_y^2$.
The overall factor of $2N_f N_c$ means the degrees of freedom for particle-antiparticle, flavors, and colors, and the spin degrees of freedom are labeled as $s=\pm$.
The overall factor of $-2$ and the factor of $-e^{-L_zk_z^{[s]}}$ are a property of the PBC on fermion fields.
See Appendix~\ref{App:Lifshitz} for a derivation of Eq.~(\ref{eq:ECas_PBC}) and the case of the antiperiodic boundary condition.
Also, for discussion with the MIT bag boundary condition, see Appendix~\ref{MIT}.

Note that Eq.~(\ref{eq:ECas_PBC}) is one of the main findings in this paper, which is an analog of the Lifshitz formula~\cite{Lifshitz:1956zz} to calculate the Casimir effect for photon fields.
When we substitute $M=q=0$ and $q=0$, Eq.~(\ref{eq:ECas_PBC}) returns to the known formulas for the massless and massive quarks, $E_{\rm Cas} = N_f N_c \times 2\pi^2 /45L_z^3$ and $E_{\rm Cas} = N_f N_c \times (2M^2 / \pi^2L_z) \sum_{l=1}^\infty K_2(l M L_z)/l^2$, respectively, where $K_2$ is the modified Bessel function.

Furthermore, we define a dimensionless quantity, which we call the {\it Casimir coefficient}, by multiplying $L_z^3$,
\begin{eqnarray}
C^{[3]}_{\rm Cas}\equiv L_z^3E_{\rm Cas}. \label{eq:CCas_cont}
\end{eqnarray}
This is a convenient quantity for checking the $L_z$ dependence of the Casimir energy.

\begin{figure*}
\includegraphics[scale=0.25]{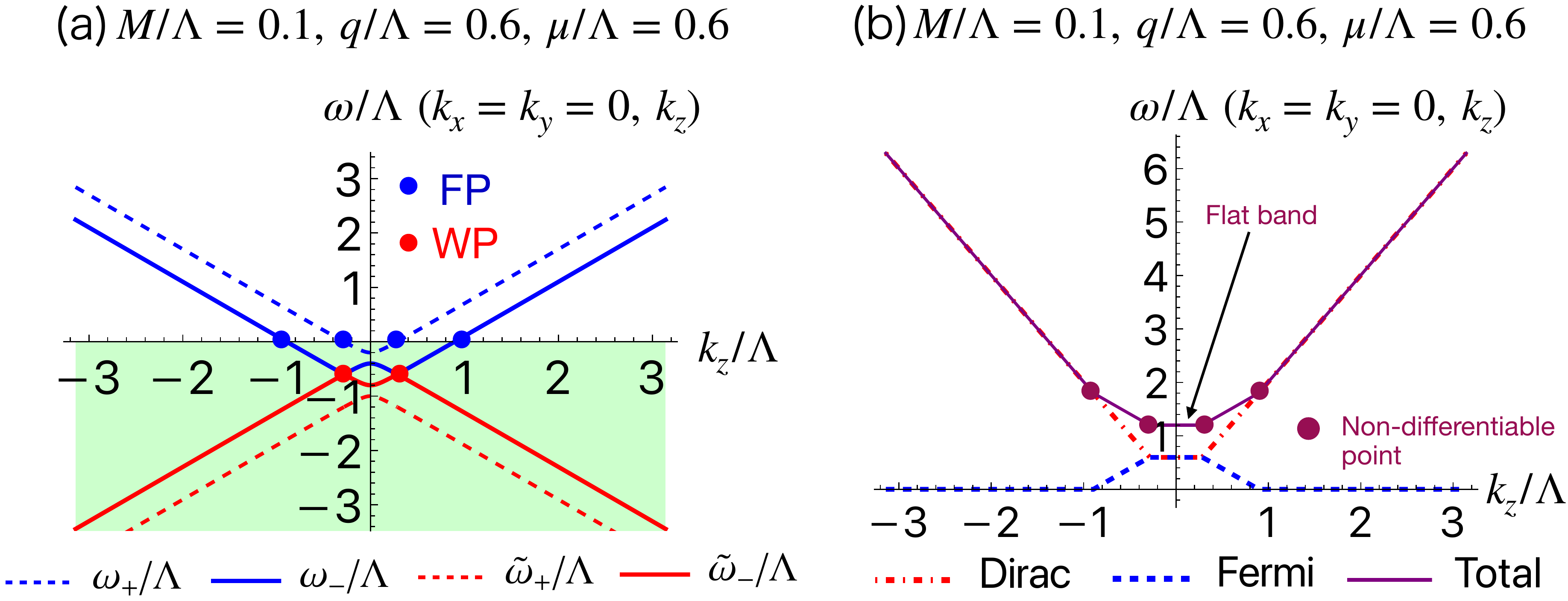}
\caption{\label{dispersion}
(a) Dispersion relations of the four eigenmodes,  Eqs.~(\ref{eq:E+-}) and (\ref{eq:tildeE+-}), at $(M/\Lambda,q/\Lambda,\mu/\Lambda)=(0.1,0.6,0,6)$.
The green region represents the Dirac and Fermi seas.
(b) Each contribution from the Dirac sea (red line) or the Fermi sea (blue line) and their sum (purple line).}
\end{figure*}

\subsection{\label{latticereg}Casimir energy with a lattice regularization}

As another approach to calculate the Casimir energy, we use lattice regularizations~\cite{Actor:1999nb,Pawellek:2013sda,Ishikawa:2020ezm,Ishikawa:2020icy,Nakayama:2022ild,Nakata:2022pen,Mandlecha:2022cll,Nakayama:2022fvh,Swingle:2022vie,Nakata:2023keh,Flores:2023whr,Nakayama:2023zvm,Beenakker:2024yhq}, which is regarded as a generalization of the original definition by Casimir~\cite{Casimir:1948dh}.
Then, the Casimir energy (per unit area) on the three-dimensional lattice with a lattice spacing $a$ is defined as
\begin{align}
E^{\rm Lat}_{\rm Cas}=&E^{\rm sum}_{0}-E^{\rm int}_{0} \label{zeropointlattice}, \\
E^{\rm sum}_0=&-\frac{N_fN_c}{a^3}\sum_{s=\pm}\int_{\rm BZ}\frac{d(ak_x)d(ak_y)}{(2\pi)^2} \notag \\
& \times\sum^{\rm BZ}_{n}\Big[\frac{1}{2}a|E^{\rm Lat}_{s,n}-\mu|+\frac{1}{2}a|E^{\rm Lat}_{s,n}+\mu|\Big], \label{eq:E0sum}\\
E^{\rm int}_{0}=&-\frac{N_fN_c}{a^3}\sum_{s=\pm}\int_{\rm BZ}\frac{d(ak_x)d(ak_y)d(ak_z)}{(2\pi)^3} \notag \label{eq:E0int}\\
& \times N_z\Big[\frac{1}{2}a|E^{\rm Lat}_{s}-\mu|+\frac{1}{2}a|E^{\rm Lat}_{s}+\mu|\Big], 
\end{align}
where the dispersion relations $E^{\rm Lat}_{\pm}$ on the lattice is obtained by replacing the original momenta $k^2_i$ ($i=x,y,z$) in Eq.~(\ref{eq:Es}) as
\begin{eqnarray}
k^2_i\rightarrow\frac{1}{a^2}(2-2\cos ak_i).
\end{eqnarray}
The first term $E^{\rm sum}_0$ in Eq.~(\ref{zeropointlattice}) is the zero-point energy in finite volume (and at finite chemical potential) with the momentum discretized by a finite thickness $L_z=aN_z$, where $N_z$ is the number of lattice cells.
The second term $E^{\rm int}_0$ in Eq.~(\ref{zeropointlattice}) is the zero-point energy with the continuous momentum in infinite volume.
The Casimir energy $E^{\rm Lat}_{\rm Cas}$ is defined as the difference between these two energies.
Also, as a result of lattice regularization, the momentum integral and the momentum sum are taken within the first Brillouin zone (BZ).
In the case of the PBC, the sum is over $n=0,1,...,N_z-1$ (or equivalently $n=1,2,...,N_z$).

Finally, we also define the dimensionless Casimir coefficient on the lattice, 
\begin{eqnarray}
C^{[3]{\rm Lat}}_{\rm Cas}\equiv L_z^3E^{\rm Lat}_{\rm Cas}=a^3N_z^3E^{\rm Lat}_{\rm Cas}. \label{eq:CCas_Lat}
\end{eqnarray}
The quantities on the lattice, $E^{\rm Lat}_{\rm Cas}$ and $C^{[3]{\rm Lat}}_{\rm Cas}$, depend on the lattice spacing $a$, but its continuum limit ($a \to 0$) should coincide with Eqs.~(\ref{eq:ECas_PBC}) and (\ref{eq:CCas_cont}) in the continuum theory (if the lattice regularization is appropriate): 
\begin{eqnarray}
C^{[3]}_{\rm Cas} = \lim_{a \to 0} C^{[3]{\rm Lat}}_{\rm Cas}.
\end{eqnarray}

\subsection{\label{Dispersion}Dispersion relations}

We remark on the dispersion relations (\ref{eq:E+-}) and (\ref{eq:tildeE+-}) for quark fields in the DCDW phase.
In Fig.~\ref{dispersion}(a), we show the four eigenmodes in the DCDW phase.
Here, the parameters are fixed as ($M/\Lambda, q/\Lambda,\mu/\Lambda,)=(0.1,0.6,0.6)$, where each quantity is dimensionless by dividing by a dimensional parameter $\Lambda$.
In this figure, the two touching points of $\omega_-$ and $\tilde{\omega}_-$ are regarded as Weyl points, and in addition, $\omega_-$ intersects the Fermi level at the Fermi points (FPs).
Furthermore, $\omega_+$ also intersects the Fermi level.
Now, since we set $q=\mu$ for simplicity, the momenta of the Weyl points of $\omega_-$ and the intersection between the Fermi level and $\omega_+$ coincide.

Figure~\ref{dispersion}(b) shows each contribution from the Dirac or Fermi sea and their sum.
The Dirac-sea and Fermi-sea contributions correspond to the absolute value of the first and second terms of the integrand in Eq.~(\ref{zeropoint}), respectively.
These dispersion relations become flat bands at low momentum and bend twice in the middle of the dispersion relation.
Such nondifferentiable points in the dispersion relation generally lead to an oscillating Casimir effect.

In the next section, we will show plots as shown in Fig.~\ref{dispersion}(b) for intuitively understanding the mechanism of the Casimir effect.

\section{\label{Results}Results}

In this section, we show the results for the Casimir coefficients in the zero-, intermediate-, and high-density regions, with the PBC (see Appendix~\ref{MIT} for the discussion with the MIT bag boundary condition).
In the following, with $\Lambda=860 \ {\rm MeV}$ and $1=\hbar c \sim 197.327 \ {\rm MeV\cdot fm}$ ($\hbar$ is the reduced Planck constant, and $c$ the speed of light), we fix the values of the order parameters, $M$ and $q$, at three $\mu$ (zero-, intermediate-, and high-density regions) as those obtained in Ref.~\cite{Tatsumi:2004dx}, where the authors applied the mean-field approach in the NJL model with the proper-time regularization.\footnote{Note that the values of order parameters, $M$ and $q$, in the NJL model depend on regularization schemes.
The Casimir energy is a quantity independent of regularization schemes.
Therefore, once $M$ and $q$ are obtained by a regularization, we can calculate the Casimir energy equivalently by any regularization (using the fixed $M$ and $q$).}

\subsection{\label{oscillating}Oscillating Casimir effect (virtual parameters)}

We will see in the following section that the oscillating Casimir effect is attributed to the flat band effect in the total dispersion relation caused by the presence of Weyl points.
Therefore, in this subsection, we show how the flat band induces the oscillating Casimir effect by using a virtual parameter set that is not a solution to the gap equation of the NJL model but gives a typical flat band. 

\begin{figure}
\includegraphics[scale=0.18]{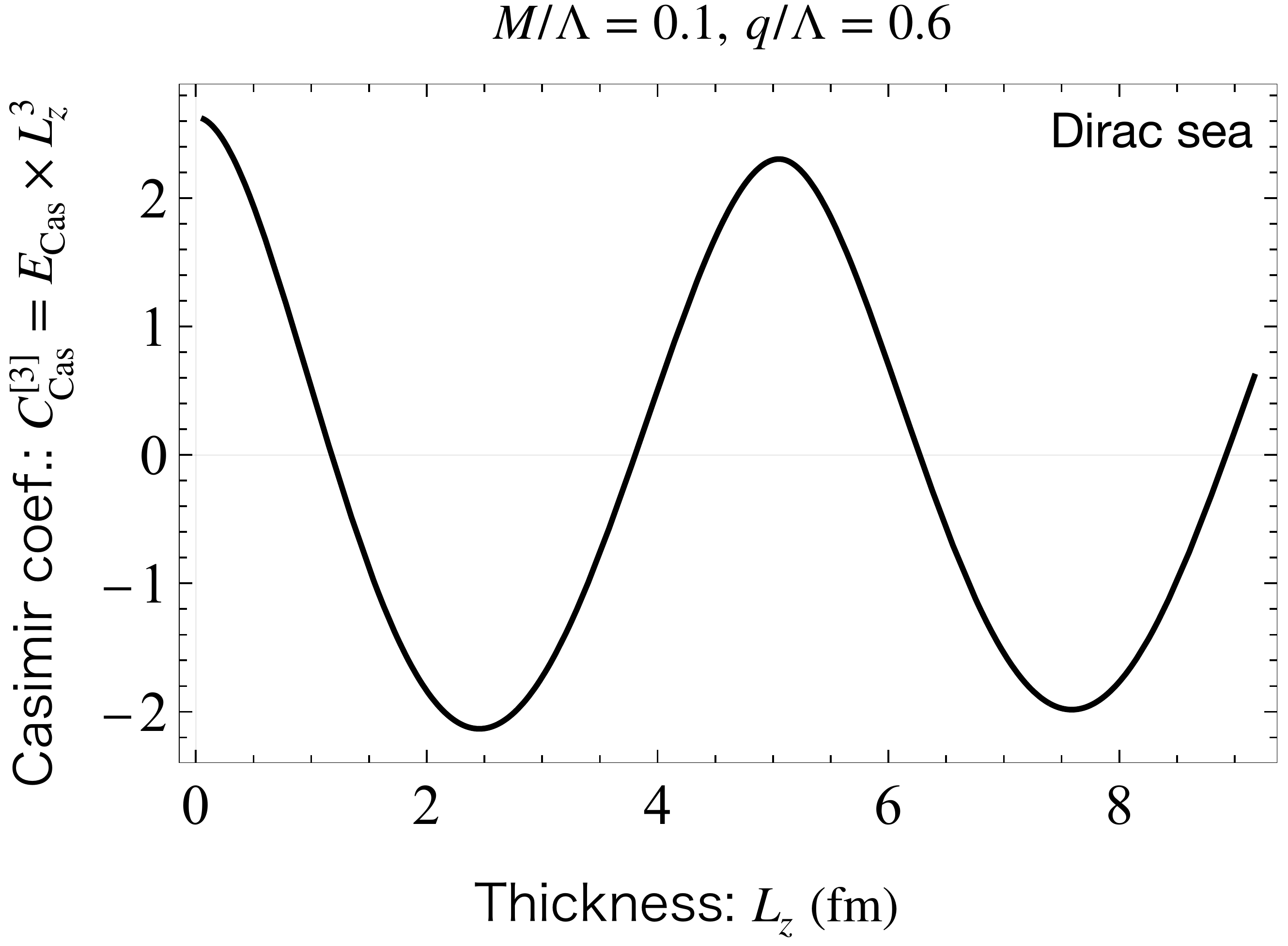}
\caption{An example of the Casimir coefficient for the oscillating Casimir effect from the Dirac-sea contribution, where we fix $(M/\Lambda,q/\Lambda)=(0.1,0.6)$ as virtual parameters.}
\label{oscillatingex} 
\end{figure}

As typical parameters, we consider $(M/\Lambda,q/\Lambda)=(0.1,0.6)$, as shown in Fig.~\ref{dispersion}.
Then, the Casimir coefficient defined as Eq.~(\ref{eq:CCas_cont}) is shown in Fig.~\ref{oscillatingex}. 
The obtained results show the oscillating Casimir effect.
The Casimir energy is obtained from the difference between the momentum integral of the dispersion relation and the infinite sum of the momentum discretized by the boundary conditions.
The oscillation of Casimir energy arises from the matching of the Weyl points and the discrete levels.

\begin{figure}
\includegraphics[scale=0.13]{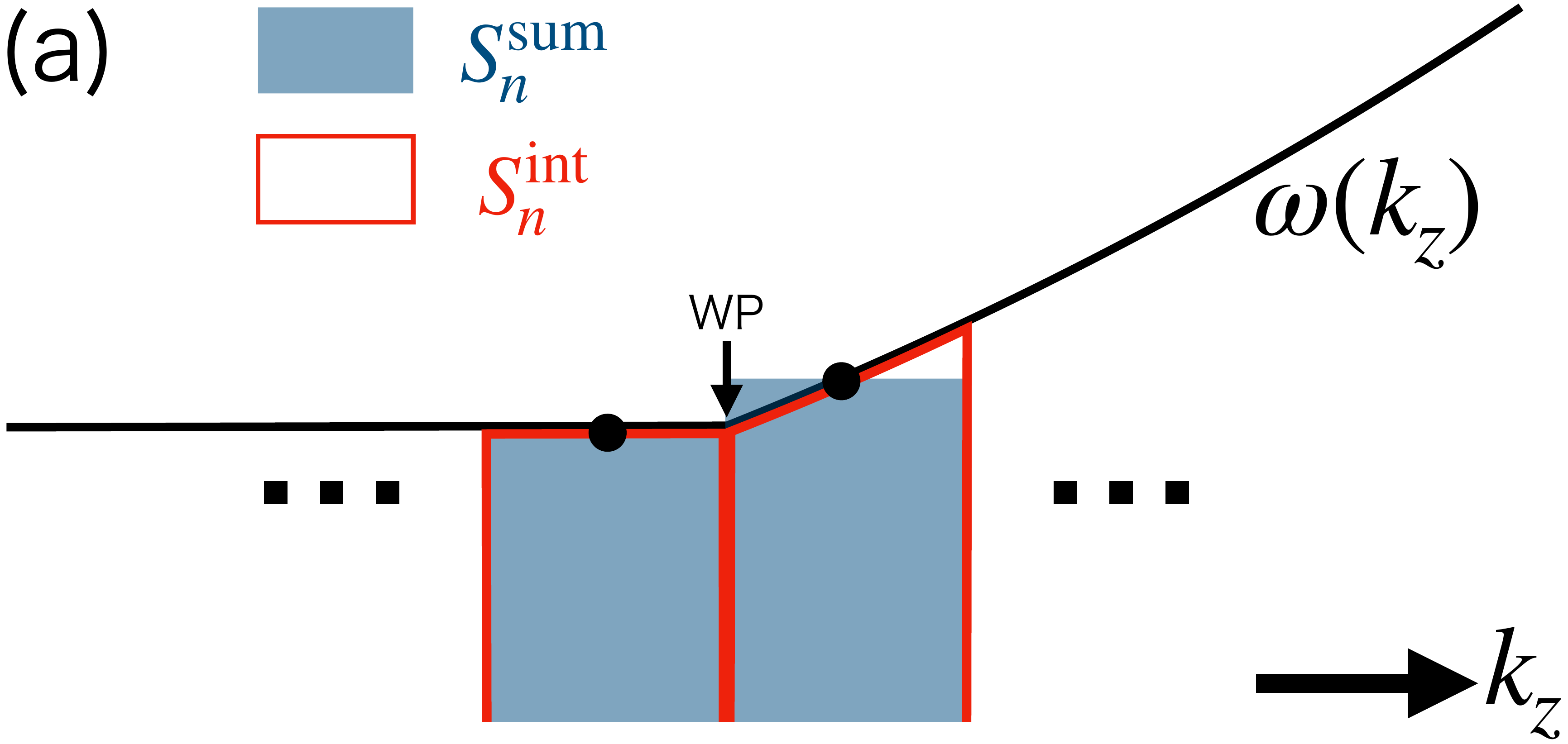}
\includegraphics[scale=0.13]{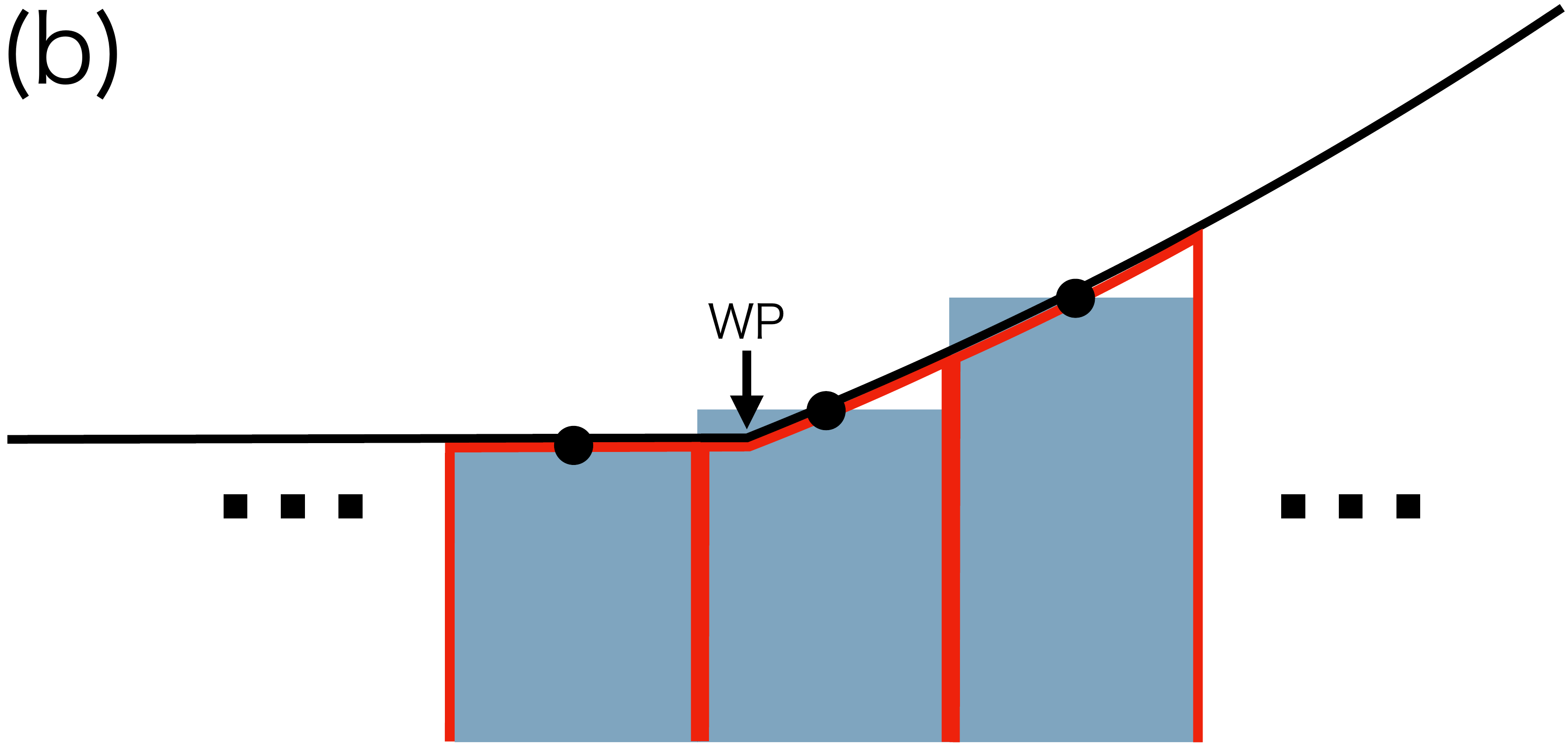}
\includegraphics[scale=0.13]{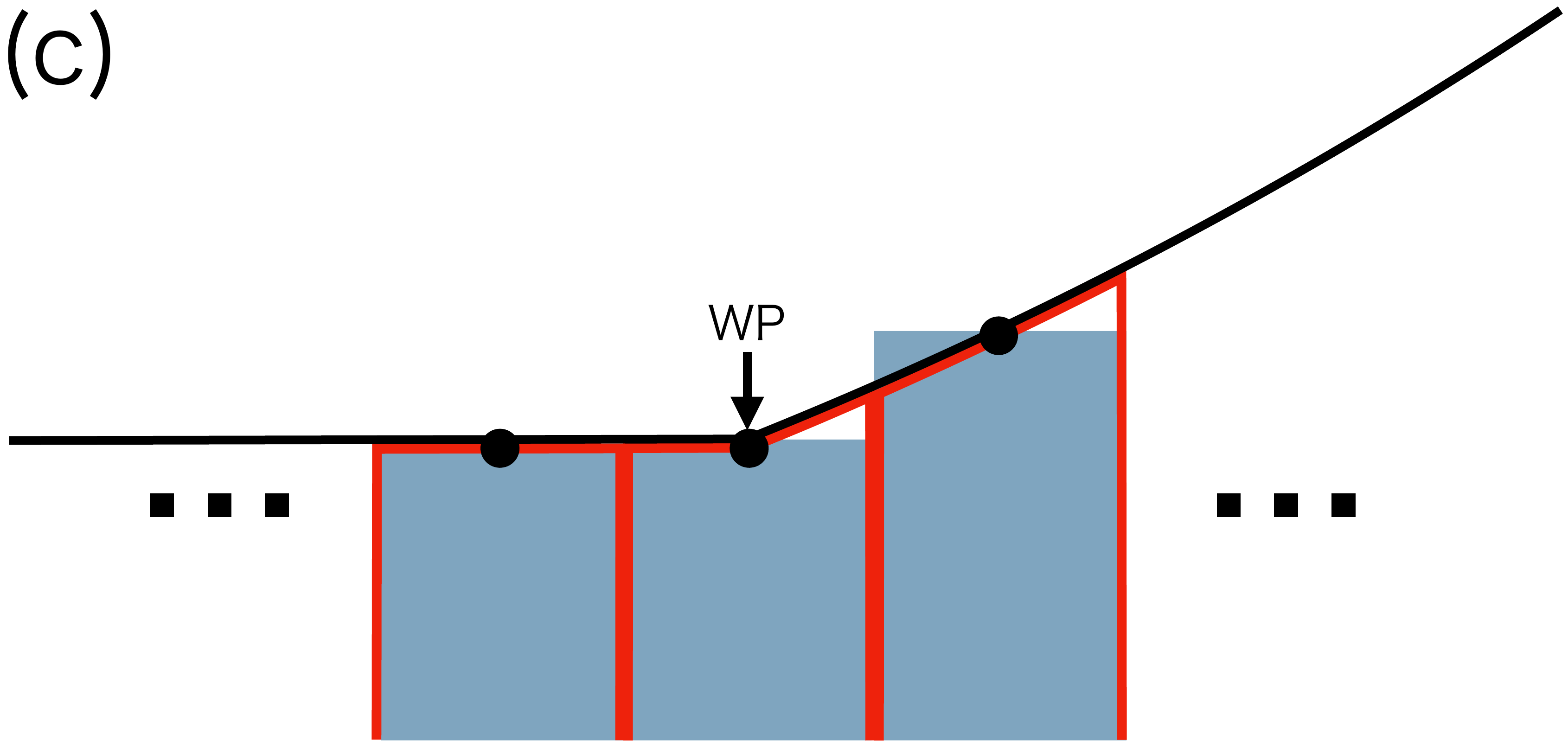}
\includegraphics[scale=0.13]{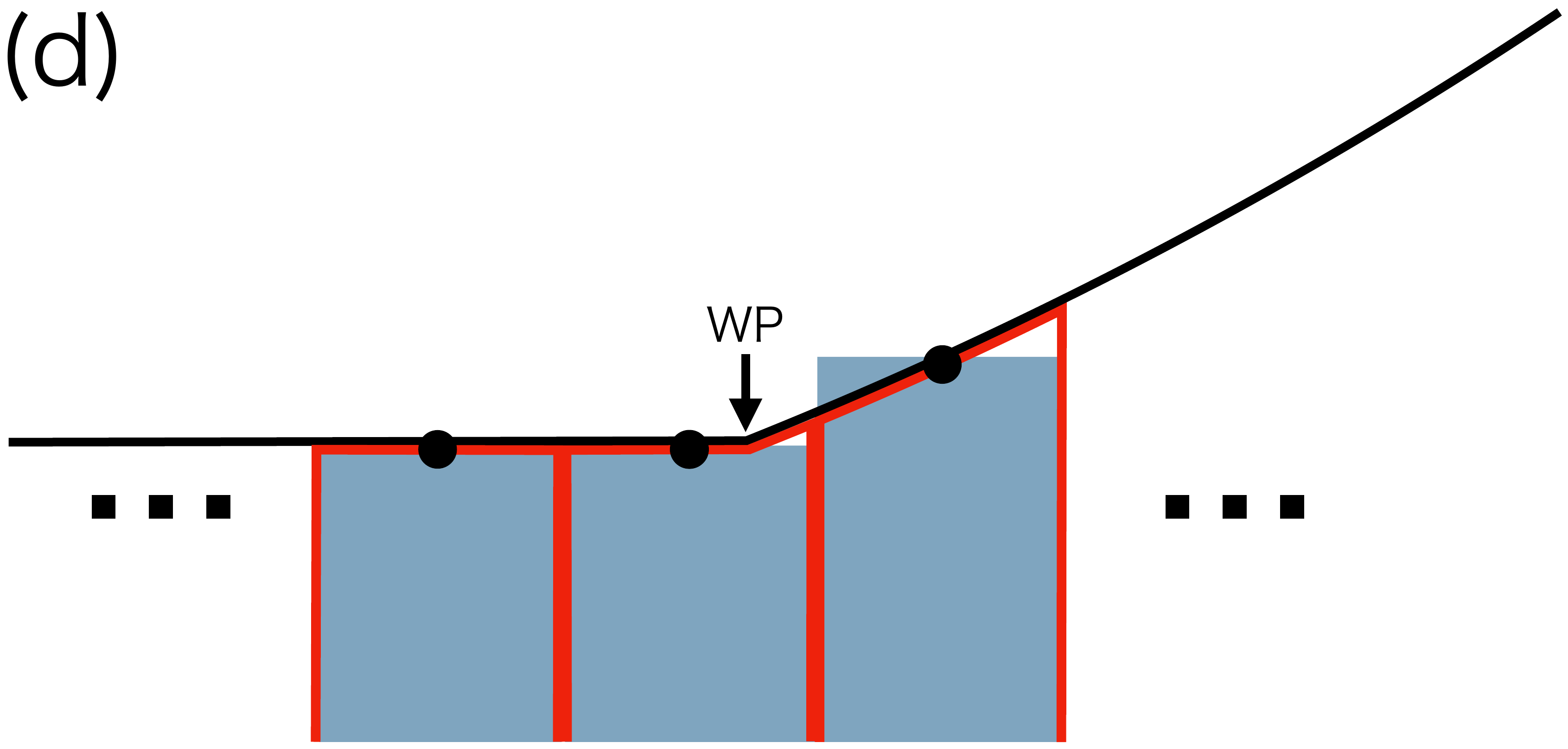}
\caption{\label{sum-integ} Graphical explanation for the oscillating Casimir effect.
The sum of the blue rectangles $S^{\rm sum}_n$ and the integral over the red areas $S^{\rm int}_n$ correspond to the zero-point energies, $E_0^{\rm sum}$ and $E_0^{\rm int}$ in finite and infinite volumes, respectively.
The Casimir energy is defined as $E_0^{\rm sum}-E_0^{\rm int}$, which corresponds to $\sum_n S^{\rm sum}_n - \sum_n S^{\rm int}_n$.
}
\end{figure}

In order to intuitively understand the mechanism of oscillations, we provide a graphical explanation.
In the following explanation, as shown in Fig.~\ref{sum-integ}, we represent $E_0^{\rm sum}$ by the sum of the blue rectangular areas $S^{\rm sum}_n$, while the corresponding $E_0^{\rm int}$ is the sum of the red areas $S^{\rm int}_n$.
For example, in the low-momentum region of a massive particle, since the dispersion relation is a downward convex function, then $S^{\rm int}_n>S^{\rm sum}_n$ is satisfied, leading to a positive contribution to $N_z(S^{\rm sum}_n-S^{\rm int}_n) \propto E^{\rm Lat}_{\rm Cas}$.\footnote{This is a simplified situation.
As a more complex case, when a dispersion relation is inverted by a Weyl point or a Fermi point, the function in the high-momentum side can be an upward convex function.}
On the other hand, in the high-momentum region, since the dispersion relation is an upward convex function due to the lattice regularization, $S^{\rm sum}_n>S^{\rm int}_n$, which leads to a negative contribution to $E^{\rm Lat}_{\rm Cas}$.
In a region that can be regarded as a linear dispersion, we have $S^{\rm sum}_n=S^{\rm int}_n$. 

As a crucial situation in this paper, in the flat-band region, it is also obvious that $S^{\rm sum}_n=S^{\rm int}_n$.
The only exception is the case containing a Weyl point, where the corresponding areas are denoted as $S^{\rm sum}_{\rm W}$ and $S^{\rm int}_{\rm W}$.
As the thickness $N_z$ increases, the position of the Weyl point relative to a rectangle changes.
The case where the endpoint of a rectangle coincides with the Weyl point is shown in Fig.~\ref{sum-integ}(a).
In this case, if the dispersion relation around the Weyl point is assumed to be linear, then $S^{\rm sum}_n=S^{\rm int}_n$ is exactly held.
Precisely speaking, it deviates somewhat from the linear dispersion and yields minor contributions as discussed in the last paragraph.
As $N_z$ increases, the size and positions of rectangles change and reach the case shown in Fig.~\ref{sum-integ}(b).
In this case, we always have $S^{\rm int}_n>S^{\rm sum}_n$, and $E^{\rm Lat}_{\rm Cas}$ increases with $N_z$.
As we continue to increase $N_z$, we eventually reach the case shown in Fig.~\ref{sum-integ}(c), where the Weyl point coincides with the midpoint of the rectangle's edge.
In this case, $S^{\rm sum}_{\rm W}$ goes under the flat band, which maximizes $E^{\rm Lat}_{\rm Cas}$.
Furthermore, as $N_z$ increases, the rectangle and the Weyl point reach the situation shown in Fig.~\ref{sum-integ}(d).
In this case, the excess of $S^{\rm int}_{\rm W}$ decreases and eventually reaches zero.
Further increasing $N_z$, the position of the Weyl point for the rectangle returns to the case of Fig.~\ref{sum-integ}(a).

Because the process from (a) to (d) occurs periodically, so that the value of $E^{\rm Lat}_{\rm Cas}$ oscillates with $L_z$.
This is the origin of the $L_z$-dependent oscillation of Casimir energy.
Then, the oscillation period (for the PBC) with respect to $L_z$ is written as~\cite{Nakayama:2022fvh}
\begin{align}
L_{z}^{\rm osc} = \frac{2\pi}{k_{\rm WP}}, \label{eq:period}
\end{align}
where $k_{\rm WP}$ is the momentum separation between the Weyl point and the origin.
For the current parameters, the period is estimated to be $L_{z}^{\rm osc} \sim 5.1$ fm with $k_{\rm WP}/\Lambda \sim 0.283$, which is consistent with the plot in Fig.~\ref{oscillatingex}.
In addition, in the $3+1$ dimensions, the minor deviation from this naive estimate is caused by the integral over the $k_x \neq 0$ and $k_y\neq 0$.

\subsection{\label{Low}Zero-density region}

First, we consider the Casimir effect in the low-chemical potential region, where we fix $\mu/\Lambda=0.46$.
In this region, we find a chiral condensed phase with $M/\Lambda=0.47$, but the DCDW phase has not yet developed $(q=0)$.

\begin{figure}
\includegraphics[scale=0.22]{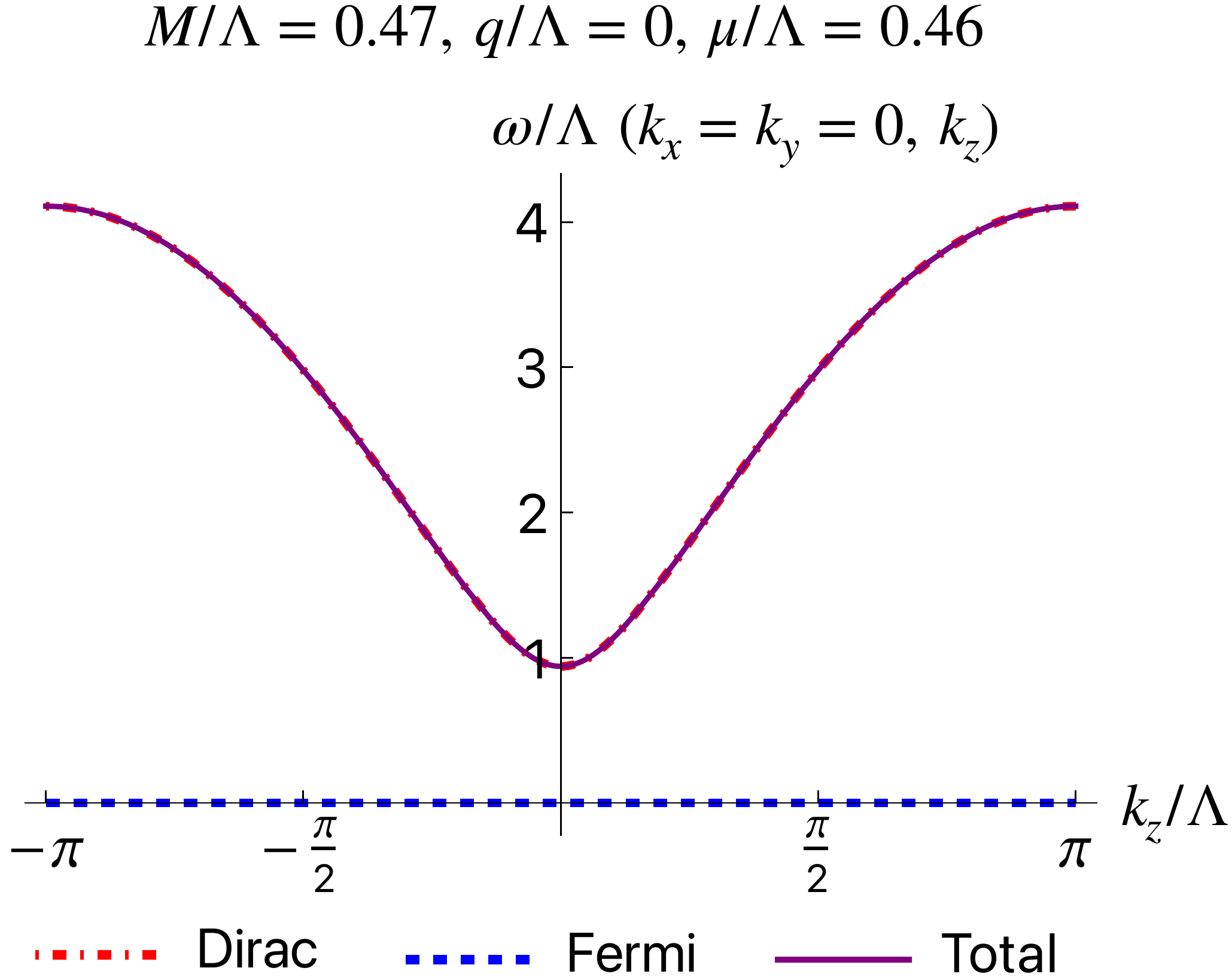}
\caption{\label{lowdispersion} Dispersion relations on the lattice with $a\Lambda=1$ in the zero-density region ($\mu/\Lambda=0.46$).}
\end{figure}

Figure~\ref{lowdispersion} shows the dispersion relation of the quark field along the momentum $k_z$ on the lattice with $a\Lambda=1$, where the red line represents the contribution from the Dirac sea, the blue line from the Fermi sea, and the purple line from their sum. 
In this case, there is no mode crossing the Fermi level, which means that the total dispersion relation has no nondifferentiable point.
Therefore, the four dispersion relations are nothing but that of a massive Dirac fermion.

\begin{figure}
\includegraphics[scale=0.18]{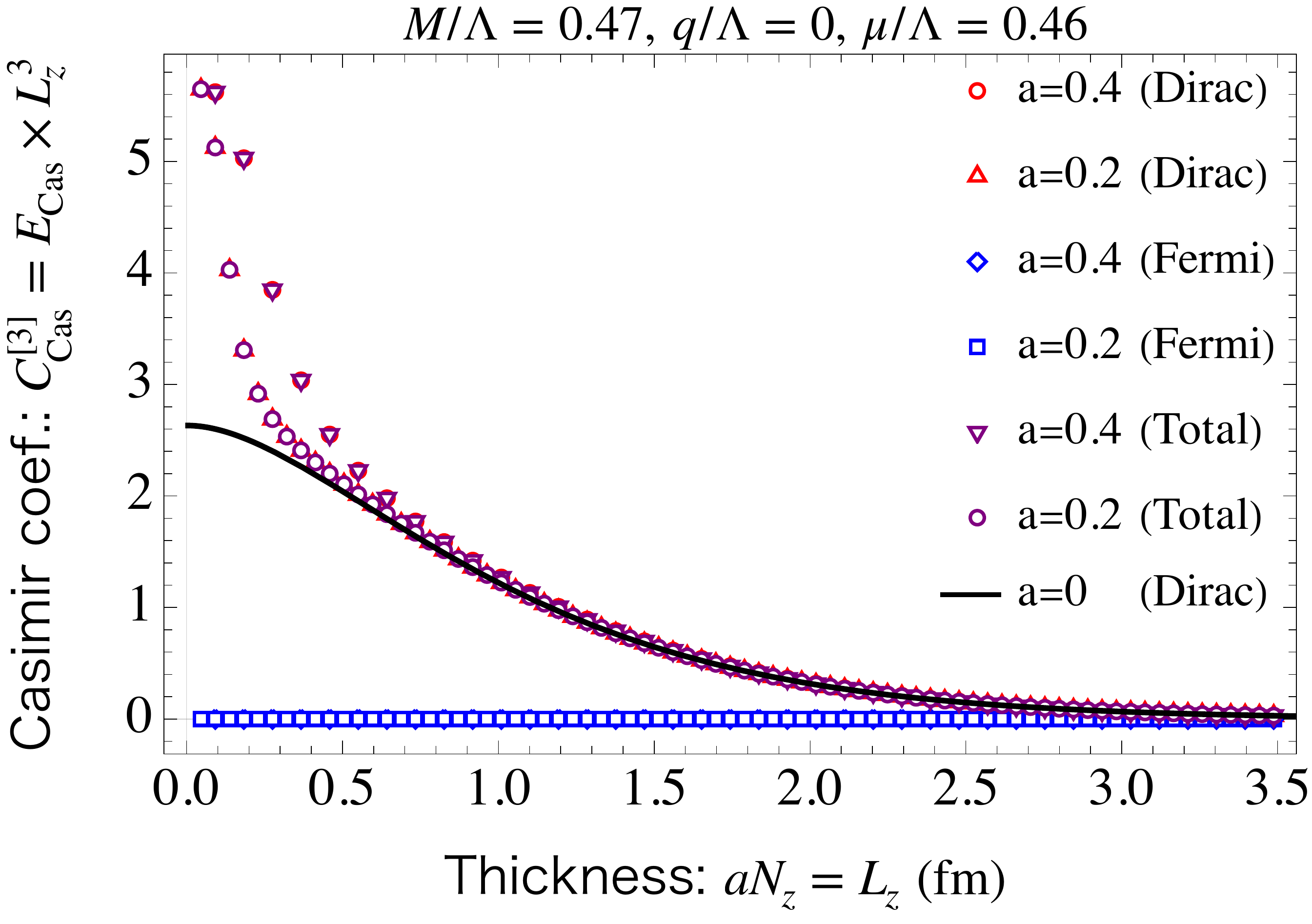}
\caption{\label{lowCCas_PBC} Casimir coefficients in the zero-density region.
We denote $a\Lambda$ as $a$ in the legends.}
\end{figure}

Figure~\ref{lowCCas_PBC} shows the results for the Casimir coefficient at $(M/\Lambda,q/\Lambda,\mu/\Lambda)=(0.47, 0,0.46)$, obtained from the two approaches. 
The resultant Casimir effect occurs only from the contribution of the Dirac sea. On the other hand, there is no contribution to the Casimir effect from the Fermi sea, as expected.
In this figure, we find that $C^{[3]}_{\rm Cas}$ approaches zero in the long-thickness region.
This behavior means that the damping of the Casimir energy $E_{\rm Cas}$ is faster than $1/L_z^3$ ($1/L_z^3$ is expected for the massless fields in the three-dimensional space).
Such a faster damping is well known for free massive fields~\cite{Bender:1976wb,Hays:1979bc,Ambjorn:1981xw}.

We present results for the two cases of lattice spacing $a\Lambda=0.2, 0.4$ in Fig.~\ref{lowCCas_PBC}.
We show analytical solutions for only the Dirac-sea part of the Casimir coefficient, obtained from the Lifshitz formula.
We observe that in the long-thickness region of about $aN_z=L_z>1.0 \ {\rm fm}$, the lattice regularization works well even for $a\Lambda=0.4$.
On the other hand, in the short-thickness region, the results for $a\Lambda=0.4$ deviate from the analytical solution due to the enhanced ultraviolet lattice cutoff effect. On the other hand, the results for $a\Lambda=0.2$ reproduce the analytical solution well up to about $aN_z=L_z=0.5 \ {\rm fm}$. This result guarantees that the results from our lattice regularization are accurate up to a quite short-thickness region. 

Finally, we comment on the physical scale of the vertical axis in Fig.~\ref{lowCCas_PBC}.
Since now we define the Casimir energy $E_\mathrm{Cas}$ as an energy per unit area, at $L_z=1$ fm in Fig.~\ref{lowCCas_PBC}, $E_\mathrm{Cas} = \hbar c \times C_\mathrm{Cas}^{[3]}/L_z^3 \sim 241$ $\mathrm{MeV}/\mathrm{fm}^2 \sim 3.86 \times 10^4$ $\mathrm{N}/\mathrm{fm}$.
As a reference, for the case of massless quarks $(M=q=0)$, $C_\mathrm{Cas}^{[3]}=N_fN_c\times2\pi^2/45 \sim 2.63$ at any $L_z$.\footnote{This $C_\mathrm{Cas}^{[3]}$ is equal to the value at $L_z =0$ in Fig.~\ref{lowCCas_PBC}.
This is because the effect of $M$ is small enough in $L_z \to 0$ due to $k_z^2 \propto 1/L_z^2 \gg M^2$.}
Then, at $L_z=1$ fm, $E_\mathrm{Cas} \sim 519$ $\mathrm{MeV}/\mathrm{fm}^2 \sim 8.32 \times 10^4$ $\mathrm{N}/\mathrm{fm}$.\footnote{This is $-192$ times larger than the well-known value $E_\mathrm{Cas} =-\hbar c \frac{\pi^2}{720L_z^3}$ of the Casimir energy for the photon fields between perfectly conducting parallel plates.
The factor of $-192$ comes from $2N_fN_c$ for the additional degrees of freedom and $16$ for the periodic boundary conditions.}
Note that the other interesting quantities are the Casimir pressure $P_\mathrm{Cas}$ and the Casimir force $F_\mathrm{Cas}$, which are defined as $P_\mathrm{Cas} \equiv F_\mathrm{Cas}/L_xL_y \equiv -\frac{d}{dL_z} E_\mathrm{Cas}$.
As a reference, for the massless quarks at $L_z=1$ fm, $P_\mathrm{Cas} \sim 2.50 \times 10^5$ $\mathrm{N}/\mathrm{fm}^2$.

\subsection{\label{Intermediate}Intermediate-density region}

\begin{figure}[b!]
\includegraphics[scale=0.22]{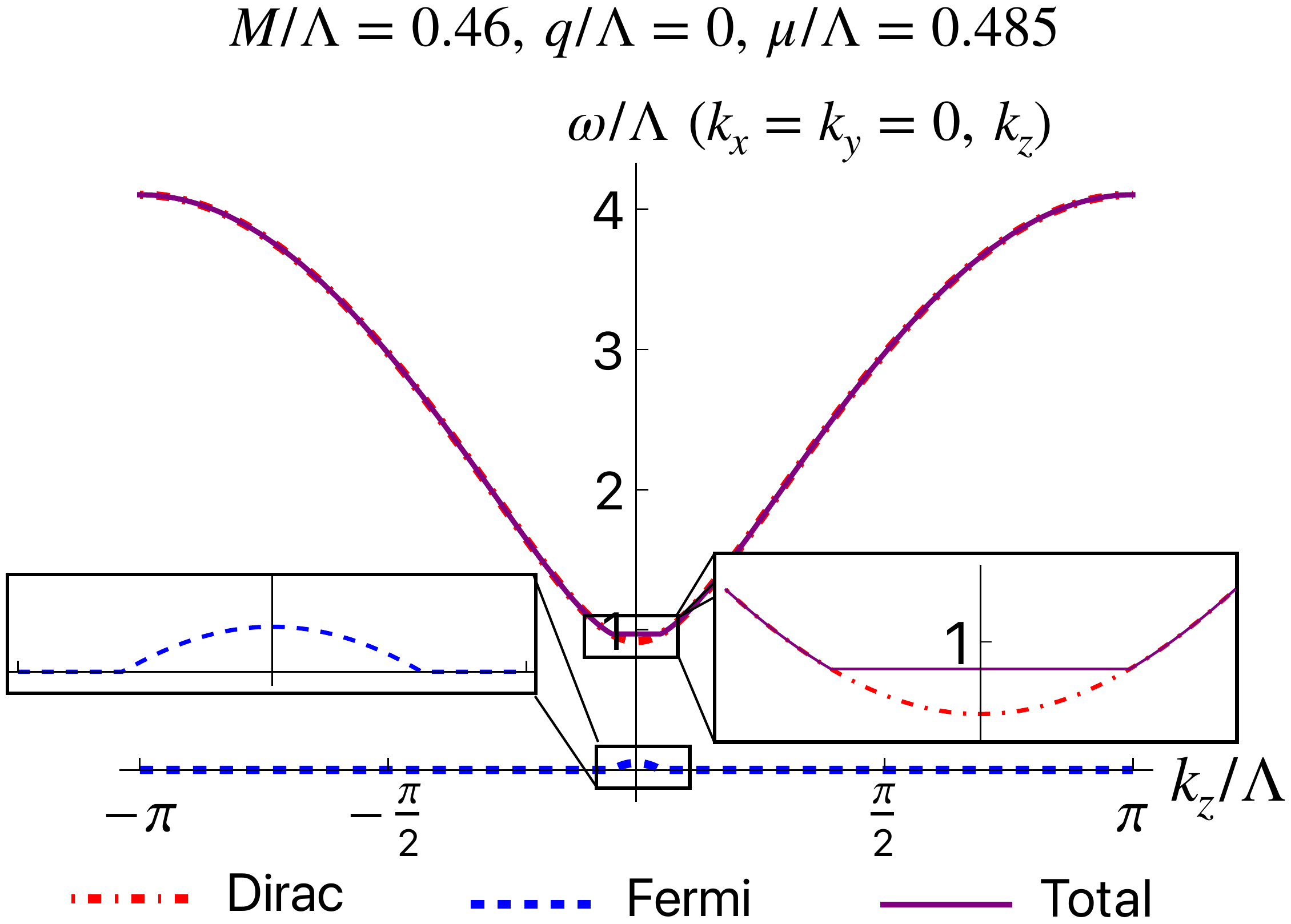}
\caption{\label{intermediatedispersion} Dispersion relations on the lattice with $a\Lambda=1$ in the intermediate-density region ($\mu/\Lambda=0.485$).}
\end{figure}
\begin{figure}[b!]
\includegraphics[scale=0.18]{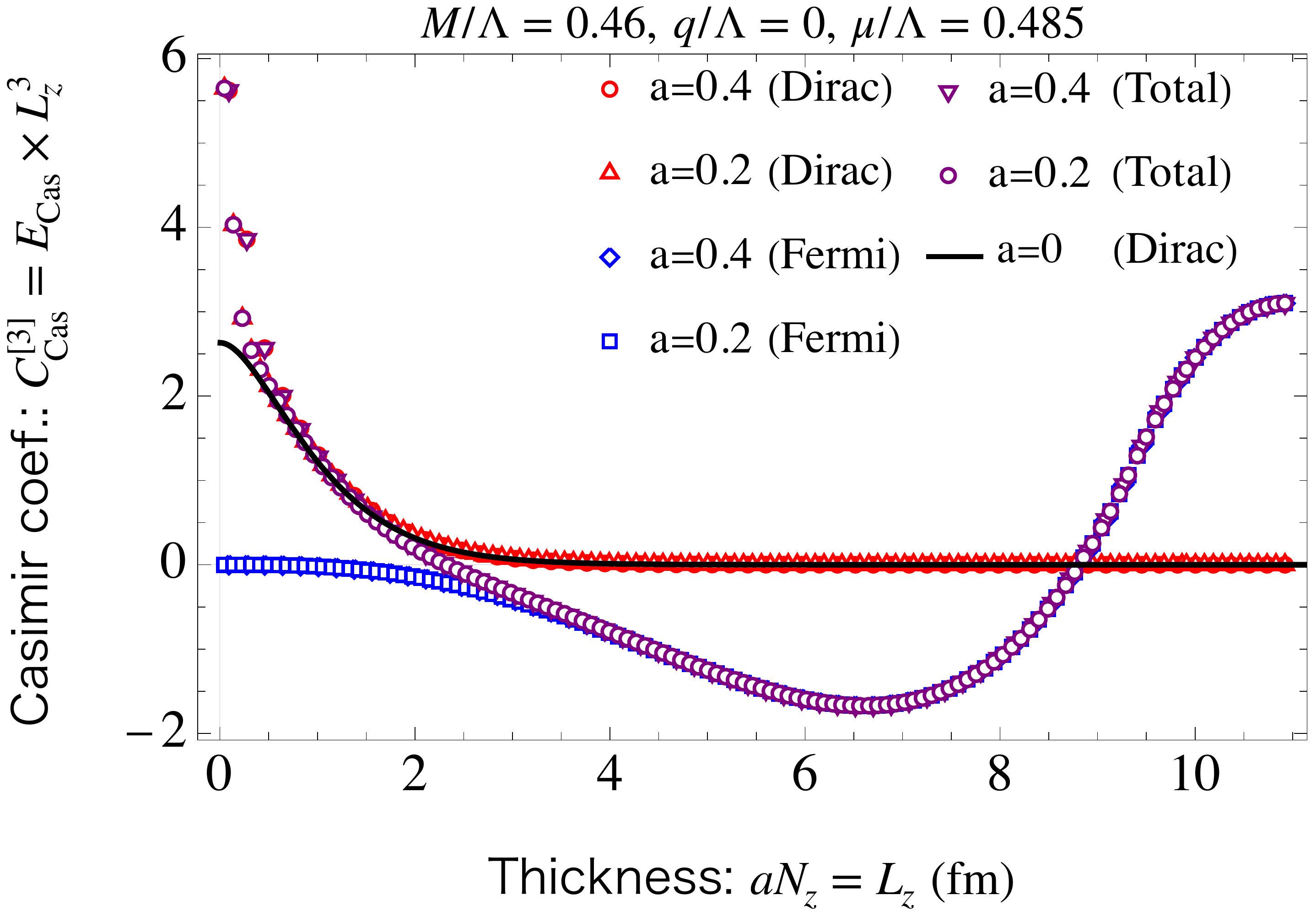}
\caption{\label{intermediateCCas_PBC} Casimir coefficients in the intermediate-density region.
We denote $a\Lambda$ as $a$ in the legends.}
\end{figure}

As the chemical potential increases, the contribution from the quark Fermi sea causes a significant change in the behavior of the Casimir effect.
Figure~\ref{intermediatedispersion} shows the dispersion relation of the quark field along the momentum $k_z$ for fixed parameters $(M/\Lambda,q/\Lambda,\mu/\Lambda)=(0.46, 0,0.485)$ and $a\Lambda=1$.
In this parameter set, the DCDW phase has not yet occurred, but the $\omega_\pm$ with low momentum is below the Fermi sea.
Then, there is an additional contribution to the Casimir effect from the Fermi sea as well as from the Dirac sea.
Furthermore, the crossing points (Fermi points) of $\omega_\pm$ and the Fermi level lead to nondifferentiable points in the total dispersion relation.
Due to this, the oscillating Casimir effect occurs, by a mechanism similar to that discussed in Sec.~\ref{oscillating}.

Figure~\ref{intermediateCCas_PBC} shows the results for the Casimir coefficient.
In the short-thickness region, we find that the Casimir coefficient is dominated by the contribution from the Dirac sea, which is qualitatively the same as the result shown in the previous Sec.~\ref{Low}.
On the other hand, in the long-thickness region, the Casimir coefficient is dominated by the contribution from the Fermi sea.
As a result, the sign of the total Casimir coefficient flips around $aN_z=L_z \sim 2.3 \ {\rm fm}$ and $ 8.8 \ {\rm fm}$.
At further long thickness, we expect an oscillation of the Casimir coefficient.
However, now we do not show the numerical results due to the limitations of calculations with sufficient accuracy.
By using Eq.~(\ref{eq:period}), the period of oscillation is estimated to be $L_z^{\rm osc}=2\pi/k_{\rm FP} \sim 9.4 \ {\rm fm}$, where $k_{\rm FP}/\Lambda \sim 0.154$ is the momentum of the Fermi point.

\subsection{\label{High}High-density region}

\begin{figure}[b!]
\includegraphics[scale=0.22]{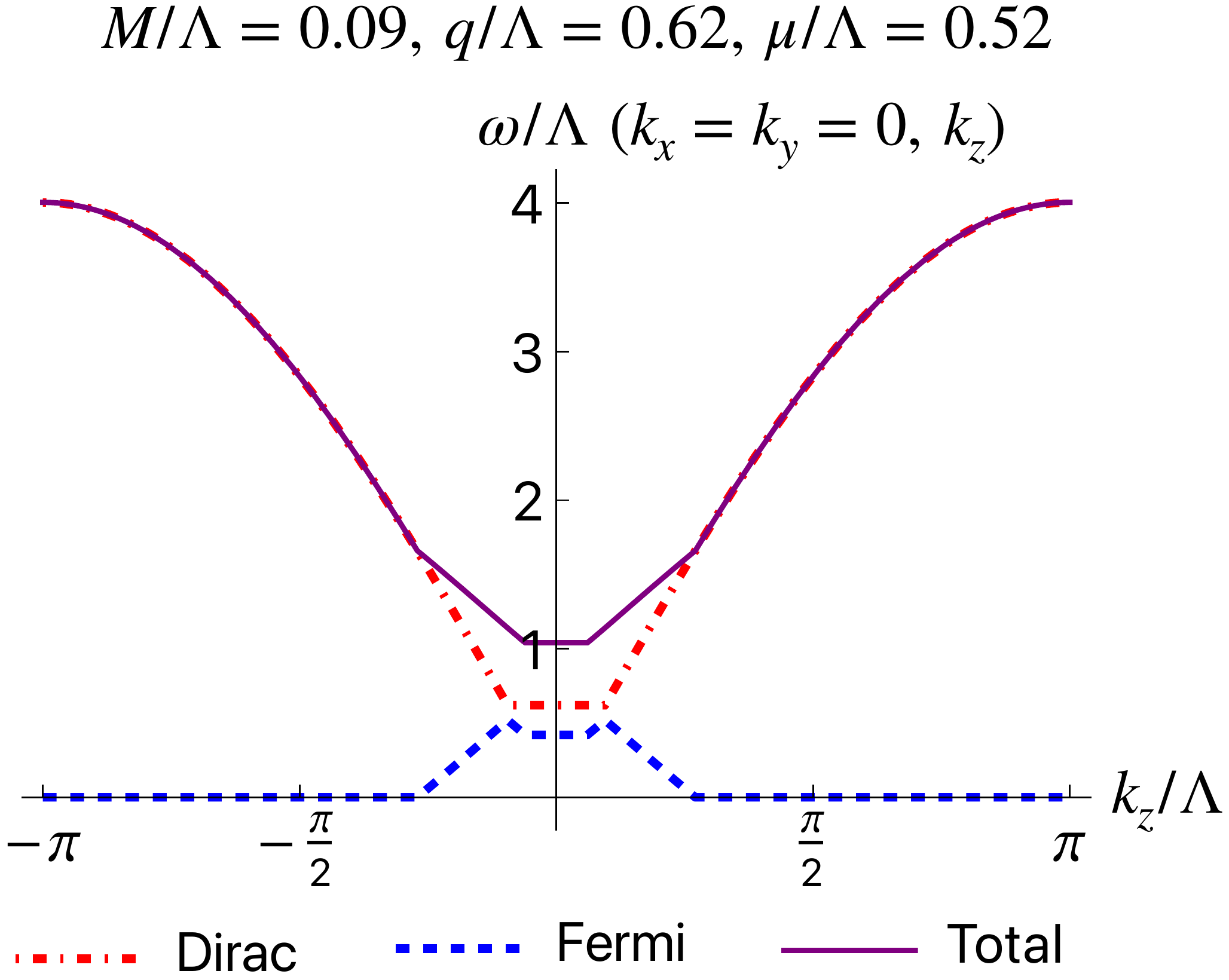}
\caption{\label{highdispersion} Dispersion relations on the lattice with $a\Lambda=1$ in the high-density region ($\mu/\Lambda=0.52$).}
\end{figure}

When the chemical potential is large enough, the DCDW phase ($q \neq 0$) is realized.
Figure~\ref{highdispersion} shows the dispersion relation of the quark fields along the momentum $k_z$ for fixed parameters $(M/\Lambda,q/\Lambda,\mu/\Lambda)=(0.09, 0.62, 0.52)$ with $a\Lambda=1$. 
In this case, the two touching points of $\omega_-$ and $\tilde{\omega}_-$ are regarded as the Weyl points.
Furthermore, the $\omega_+$ and $\omega_-$ cross the Fermi level.
Thus, there are six nondifferentiable points.
In the low-momentum region, the total dispersion relation becomes a flat band due to the cancellations between $\omega_-$ and $\tilde{\omega}_-$ and between $\omega_+$ and $\tilde{\omega}_+$, and the edges of the flat band of the total dispersion relation are nondifferentiable.
Furthermore, due to the four Fermi points, the total dispersion relation is nondifferentiable at the four momenta.

\begin{figure}[t!]
\includegraphics[scale=0.18]{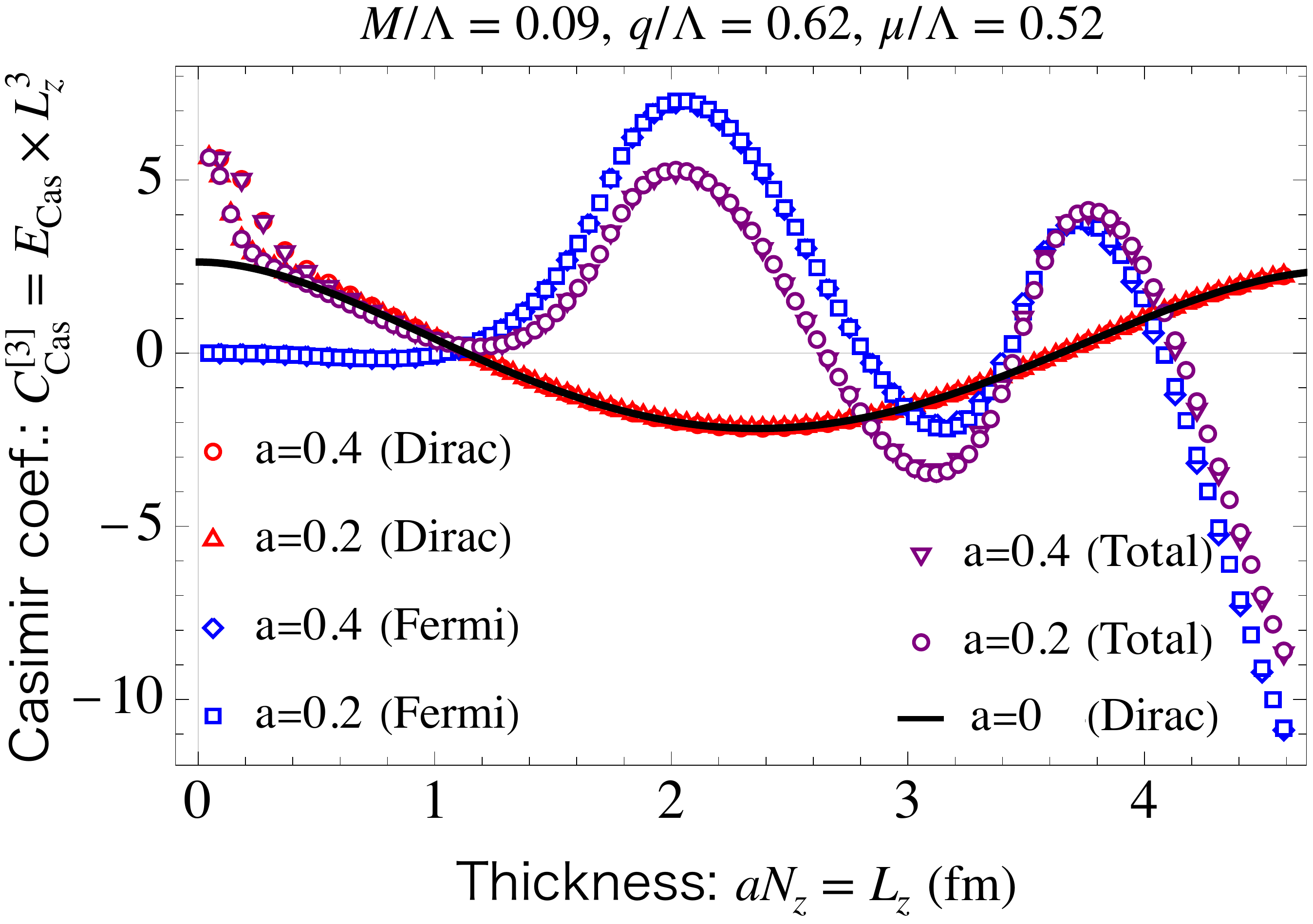}
\caption{\label{highCCas_PBC} Casimir coefficients in the high-density region.
We denote $a\Lambda$ as $a$ in the legends.}
\end{figure}

Figure~\ref{highCCas_PBC} shows the results for the Casimir coefficient. 
For the Dirac-sea effect, the Casimir effect shows oscillating behavior, as expected from the Weyl points.
Furthermore, we see that the Fermi-sea part of Casimir coefficient oscillates with higher frequency than that from the Dirac sea.
The frequencies of the oscillations are determined by the positions of the nondifferentiable points on the dispersion relation.
By using Eq.~(\ref{eq:period}), the oscillation period is estimated to be $L_z^{\rm osc}=2\pi/k_{\rm WP} \sim 4.8$ fm with $k_{\rm WP}/\Lambda \sim 0.298$ for the Dirac-sea part.
For the Fermi-sea part of $\omega_-$, the period is determined by $L_z^{\rm osc}=2\pi/k_{\rm FP} \sim 1.7$ fm with $k_{\rm FP}/\Lambda \sim 0.851$ in addition to $L_z^{\rm osc}=2\pi/k_{\rm WP} \sim 4.8$ fm.
For the Fermi-sea part of $\omega_+$, $L_z^{\rm osc}=2\pi/k_{\rm FP} \sim 7.6$ fm with $k_{\rm FP}/\Lambda \sim 0.190$.
In fact, the Dirac-sea and Fermi-sea contributions shown in Fig.~\ref{highCCas_PBC} are dominated by the periods of $L_z^{\rm osc}\sim 4.8$ fm and $L_z^{\rm osc} \sim 1.7$ fm, respectively, and the total result is their superposition.

\section{Conclusion and outlook} \label{Conclusions}
In this paper, we have discussed the Casimir effect from quark fields in some phases of quark matter.
In particular, we consider the QCD counterpart of a new type of Casimir effect i.e., the oscillating Casimir effect.
We have classified the typical behaviors of the Casimir effect by focusing on the three types of density regions:
\begin{enumerate}
\renewcommand{\labelenumii}{\arabic{enumii}}
\item When the quark chemical potential is small enough, the quark-number density is zero.
Then, the Casimir effect for massive quark fields occurs, where the damping of the Casimir energy with increasing thickness is characterized by the dynamical masses of quarks.
\item As the quark chemical potential increases, the quark number density becomes nonzero.
In this region, the contribution from the quark Fermi sea effect leads to the oscillating Casimir effect.
\item Further increasing the chemical potential eventually leads to a phase transition into the DCDW phase.
In this region, in addition to the Fermi sea effect, the Weyl-point structure of quark fields in the DCDW phase results in the oscillating Casimir effect.
The total Casimir energy behaves as the superposition of two types of oscillations.
\end{enumerate}

To clarify terminology, one of our main findings may be called the {\it DCDW-induced oscillating Casimir effect}, which is a kind of the {\it WP-induced oscillating Casimir effect} predicted in Ref.~\cite{Nakayama:2022fvh}. 
On the other hand, the {\it FP-induced oscillating Casimir effect} would be a common term for general fermionic systems at finite density in finite volume.

In the following, we list some potential future studies.
\begin{enumerate}
\renewcommand{\labelenumi}{(\roman{enumi})}
\item {\it Solving gap equation in finite volume}---In this work, we input the values of order parameters obtained from the gap equation in infinite volume, in order to understand the typical behavior of the Casimir effect for typical dispersion relations.
As a more precise analysis, one can calculate the order parameters by minimizing the thermodynamic or effective potential in finite volume, and then one can understand the self-consistent relationship between the volume-dependent order parameters and the Casimir effect.
This type of study would help elucidate the phase diagram of interacting fermions in finite volume (e.g., see Refs.~\cite{Kim:1987db,Braun:2004yk,Abreu:2006pt,Ebert:2010eq,Flachi:2013bc,Ishikawa:2018yey,Ishikawa:2019dcn,Inagaki:2021yhi}) from the viewpoint of the Casimir effect.\footnote{As shown in Ref.~\cite{Ebert:2010eq}, the phase diagram in the NJL model at finite $(\mu,L_z)$ is complicated, where the phase boundary between low-density and high-density phases oscillates as a function of $ L_z$.
Therefore, for a more realistic prediction of the $L_z$ dependence of the Casimir energy, such $L_z$-dependent phase transitions should be considered.
In addition, a question is whether or not the DCDW phase survives on the phase diagram.
As a preliminary result, we solved a gap equation at finite $(\mu,L_z)$ and confirmed that in fact the DCDW phase appears in a parameter region~\cite{Fujii:202xx_pre}.
In this region, our prediction of the Casimir effect is qualitatively reliable.}
\item {\it Lattice simulations}---Our predictions will be examined by future lattice NJL (and also QCD) simulations.
Since Monte Carlo simulations at large quark chemical potential suffer from the sign-problem, sign-problem-free techniques should be applied.
In particular, numerical simulations of Casimir effects for lattice gauge fields have been vigorously studied, such as the U(1) gauge~\cite{Pavlovsky:2009kg,Pavlovsky:2010zza,Pavlovsky:2011qt}, the compact U(1) gauge~\cite{Pavlovsky:2009mt,Chernodub:2016owp,Chernodub:2017mhi,Chernodub:2017gwe,Chernodub:2022izt}, the SU(2) gauge~\cite{Chernodub:2018pmt,Chernodub:2018aix}, and the SU(3) gauge fields~\cite{Kitazawa:2019otp,Chernodub:2023dok}, so that the interplay between quark and gluon dynamics would be interesting.
\item {\it Finite temperature}---At finite temperature, thermal fluctuations (as well as quantum fluctuations) also induce the Casimir effect, which is the so-called thermal Casimir effect well established theoretically~\cite{Lifshitz:1956zz} and experimentally~\cite{Sushkov:2010cv}.
The Casimir energy at finite temperature can be calculated by using a straightforward modification of Eq.~(\ref{eq:ECas_PBC}) or Eq.~(\ref{zeropointlattice}).
Although the DCDW phase disappears at sufficiently high temperature, in the low-temperature region the oscillating Casimir effect can be realized.
\item {\it Axion electrodynamics}---In this work, we consider only the Casimir effect for quark fields, which is induced by boundary conditions on quark fields.
The dynamics of photons in the DCDW phase may be described by modified Maxell equations~\cite{Tatsumi:2014wka,Ferrer:2016toh,Tatsumi:2018ifx} which is the so-called axion electrodynamics~\cite{Sikivie:1983ip,Wilczek:1987mv}.
The Casimir energy for such modified photon fields can exhibit a sign-flipping behavior~\cite{Fukushima:2019sjn,Brevik:2021ivj,Canfora:2022xcx,Oosthuyse:2023mbs,Favitta:2023hlx,Nakayama:2023zvm} when boundary conditions on photon fields are imposed.
It will be interesting whether or not such a photonic Casimir effect occurs also in the DCDW phase. 
\item {\it Models with nucleons}---In this work, we consider the Casimir effect for quark degrees of freedom based on the NJL model.
Apart from the NJL model, in the low-density phase of QCD, relevant degrees of freedom are hadrons, specifically nucleons.
Therefore, it will be important to examine the consistency between the quark Casimir effect in the NJL model and the nucleon Casimir effect in an effective model such as the nucleon linear sigma model which can also investigate the DCDW phase.
\item {\it Other inhomogeneous chiral phases}---We have shown the WP-induced oscillating Casimir effect in the DCDW phase, which is particular to systems with momentum-dependent Weyl points.
It would be interesting what types of Casimir effects can occur in other inhomogeneous chiral phases.
For example, the real-kink crystal phase is known as a possible ground state in the 3+1 dimension NJL model~\cite{Nickel:2009wj}.
In this phase, because $k_z$ is not a conserved quantity, our approach formulated in spatial-momentum space cannot be applied, but the Casimir energy should be calculated by approaches formulated with discrete eigenenergies.
Furthermore, apart from the dense quark matter, a magnetic field can also induce the DCDW phase~\cite{Frolov:2010wn}, which is the so-called magnetic dual chiral density wave (MDCDW) (see Ref.~\cite{Ferrer:2021mpq} for a review).
The Casimir effect in such a phase is also interesting.
\item {\it Lower-dimensional models of QCD}---The WP-induced oscillating Casimir effect can occur also in lower spatial dimensions (i.e., $1+1$ or $2+1$ dimensions)~\cite{Nakayama:2022fvh}.
Therefore, investigation of the Casimir effect in possible inhomogeneous phases of low-dimensional QCD-like models, such as the Gross-Neveu model~\cite{Thies:2003br,Thies:2003kk,Lenz:2020bxk}, the chiral Gross-Neveu (NJL$_2$) model~\cite{Schon:2000he}, and the 't Hooft model (QCD$_2$)~\cite{Schon:2000he,Kojo:2011fh,Hayata:2023pkw}, will be also important.
\end{enumerate}

\section*{ACKNOWLEDGMENTS}
The authors thank Tsutomu Ishikawa for fruitful discussions.
This work was supported by the Japan Society for the Promotion of Science (JSPS) KAKENHI (Grant No. JP20K14476).

\appendix

\section{Lifshitz formula\label{App:Lifshitz}}

In this Appendix, we derive the Lifshiz formula~(\ref{eq:ECas_PBC}) for the dispersion relations in the DCDW phase.
To calculate Casimir energy at zero temperature and zero chemical potential, we need to obtain the infinite sum of eigenvalues of quarks, 
\begin{align}
\begin{aligned}  
    &\sum_{s=\pm}\sum^\infty_{n=-\infty}\frac{\omega_{s,n}^\mathrm{PBC}}{2}
    =\sum_{s=\pm}{\sum^\infty_{n=0}}^\prime \omega_{s,n}^\mathrm{PBC}, \\
    &\sum_{s=\pm}\sum^\infty_{n=-\infty}\frac{\omega_{s,n}^\mathrm{APBC}}{2}
    =\sum_{s=\pm}{\sum^\infty_{n=0}} \omega_{s,n}^\mathrm{APBC},
\end{aligned}
    \label{eq:infsum}
\end{align}
where the momentum in the $z$ direction is discretized as $k_z=2n\pi/L_z$ and $k_z=(2n+1)\pi/L_z$ under the periodic boundary conditions (PBCs) and the antiperiodic boundary conditions (APBCs), respectively.
The prime in the sum for the PBC means that the factor $1/2$ is multiplied only for $n=0$: $\sum_n^\prime \omega_n =\sum_n \omega_n - \omega_{n=0}/2$.

The discretized $k_z$ is given as the zero point of the function 
\begin{align}
    &\Delta_\pm^\mathrm{PBC} (\omega)=1-e^{-ik^{[\pm]}_zL_z}, \\
    &\Delta_\pm^\mathrm{APBC} (\omega)=1+e^{-ik^{[\pm]}_zL_z}, \\
    & ik^{[\pm]}_z \equiv \tilde{k}^{[\pm]}_z =\sqrt{k^2_\perp +M^2-\omega^2-\frac{q^2}{4}\mp iq\sqrt{k^2_\perp-\omega^2}}, \notag
\end{align}
where the imaginary momentum $ik^{[\pm]}_z$ is obtained by solving the DCDW dispersion relation $\omega_s=\sqrt{k_\perp^2+(\sqrt{k^2_z+M^2}+sq/2)^2}$ with $s=\pm 1$.
Since the function $\Delta_s(\omega)$ is a meromorphic function in a closed path $C$ on the complex $\omega$ plane with the mentioned zero point but no pole, by using the generalized argument principle~\cite{Kampen:1968,Schram:1973,Bordag:2001qi}, the infinite sum~(\ref{eq:infsum}) can be calculated from the following contour integral:
\begin{align}
    &\sum_{s=\pm}\sum^\infty_{n=-\infty}\frac{\omega_{s,n}}{2}=\sum_{s=\pm}{\sum^\infty_{n=0}}^\prime \omega_{s,n}=\sum_{s=\pm}\frac{1}{2\pi i}\oint_C \omega d\ln\Delta_s(\omega) \notag \\
    &=\sum_{s=\pm}\frac{1}{2\pi i}\Big(\int^{-i\infty}_{i\infty}\omega d\ln\Delta_s(\omega)+\int_{C_+}\omega d\ln\Delta_s(\omega)\Big),
\end{align}
where $C_+$ is the counterclockwise integral on a semicircle with an infinite radius centered at the origin, in the right half of the complex $\omega$ plane.
More precisely, when at $M=q=k_\perp=0$ under the PBCs, the zero mode exists at the origin, and then the contour needs to avoid the origin by considering an infinitesimal semicircular path centered at the origin.
By introducing the imaginary frequency $\omega \equiv i\xi$ with the imaginary part $\xi$, we obtain 
\begin{align}
    &\sum_{s=\pm}\sum^\infty_{n=-\infty}\frac{\omega_{s,n}}{2} \notag \\
    &=\sum_{s=\pm}\frac{1}{2\pi}\Big(\int^{-\infty}_{\infty}\xi d\ln\Delta_s(i\xi)+\int_{C_+}\xi d\ln\Delta_s(i\xi)\Big). \label{Arg}
\end{align}
The second term of Eq.~\eqref{Arg} on a circle with an infinite radius vanishes in the limit of $\xi\to \infty$:
\begin{align}
    \xi d\ln\Delta_s(i\xi)=\frac{L_z\xi^2e^{-\xi L_z}}{(1-e^{-\xi L_z})\xi}\xrightarrow[\xi\rightarrow\infty]{}0.
\end{align}
Using integration by parts, the first term of Eq.~\eqref{Arg} is reduced to
\begin{align}
    &\int^{-\infty}_{\infty}\xi d\ln\Delta_s(i\xi)=\big[\xi\ln\Delta(i\xi)\big]^{-\infty}_\infty+\int^\infty_{-\infty}d\xi\ln\Delta_s(i\xi) \notag \\
    &=0+2\int^\infty_0d\xi\ln\Delta_s(i\xi).
\end{align}
In the last form, we used $\xi \ln (1\pm e^{-\xi}) \to 0$ in the limit of $\xi\to\infty$ and the fact that the integrand of the second term is an even function of $\xi$.

By taking into account the $k_\perp$ integral and multiplying the factors $2N_fN_c$ for the particle-antiparticle, flavors, and colors, we obtain the Lifshiz formula~(\ref{eq:ECas_PBC}) for the PBC.
Similarly, the Lifshitz formula for the APBC is
\begin{align}
E_{\rm Cas}^{\rm APBC}
&=-4N_f N_c\sum_{s=\pm}\int_0 ^\infty\frac{d\xi}{2\pi}\int\frac{dk_xdk_y}{(2\pi)^2}\ln\big[1+e^{-L_z\tilde{k}_z^{[s]}}\big],\notag\\
\tilde{k}^{[\pm]}_z&=\sqrt{k_\perp^2+M^2+\xi^2-\frac{q^2}{4}\mp iq\sqrt{k_\perp^2+\xi^2}}. \label{eq:ECas_APBC}
\end{align}
The overall $-2$ and the factor of $+e^{-L_z\tilde{k}_z^{[s]}}$ is a property of the APBC on fermion fields.
When we substitute $M=q=0$ and $q=0$, Eq.~(\ref{eq:ECas_APBC}) returns to the known formulas for the massless and massive quarks, $E_{\rm Cas}^{\rm APBC} = -N_f N_c \times 7\pi^2 /180L_z^3$ and $E_{\rm Cas}^{\rm APBC} = N_f N_c \times (-2M^2 / \pi^2L_z) \sum_{l=1}^\infty (-1)^{l+1} K_2(l M L_z)/l^2$, respectively.
The discrete momentum, $k_z\to (2n+1)\pi/L_z$, for the APBCs is shifted by a half period from $k_z\to 2n\pi/L_z$ for the PBCs, and hence the sign of the Casimir energy is flipped and also its absolute value changes slightly.

\section{\label{MIT}Results with MIT bag boundary}

In the main text, we applied the PBC to consider the Casimir effect.
In this Appendix, we consider the case with the MIT bag boundary condition, which is more realistic when we consider quark matter in nature or experiments.

The MIT bag boundary condition~\cite{Chodos:1974je} requires that the quark-current flux toward normal to the boundary is zero.
For the massless Dirac field on the MIT bag boundaries, one can obtain the discretized momentum $k_z\to (n+1/2)\pi/L_z$ as an analytic solution satisfying the Dirac equation.
On the other hand, for the massive case, one cannot obtain an analytic form of momentum, so that one should apply an alternative approach with no analytic form of $k_z$ (e.g., Refs.~\cite{Mamaev:1980jn,Santos:2002jp,Elizalde:2002wg}).
In this Appendix, as an approximate estimate, we show the results using the massless solution $k_z\to (n+1/2)\pi/L_z$, which may be called ``MIT bag" boundary conditions.

\begin{figure}[t!]
\includegraphics[scale=0.15]{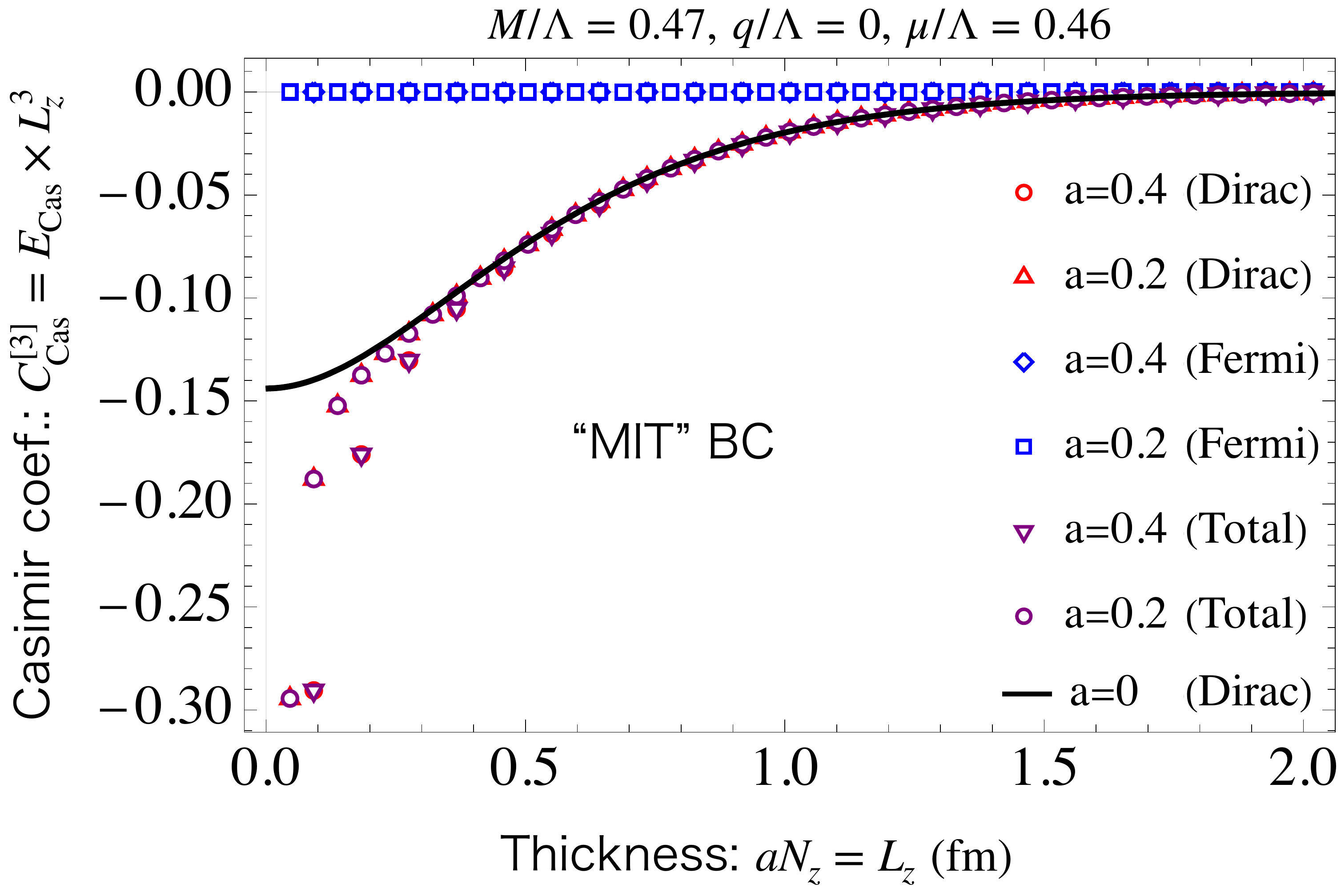}
\includegraphics[scale=0.15]{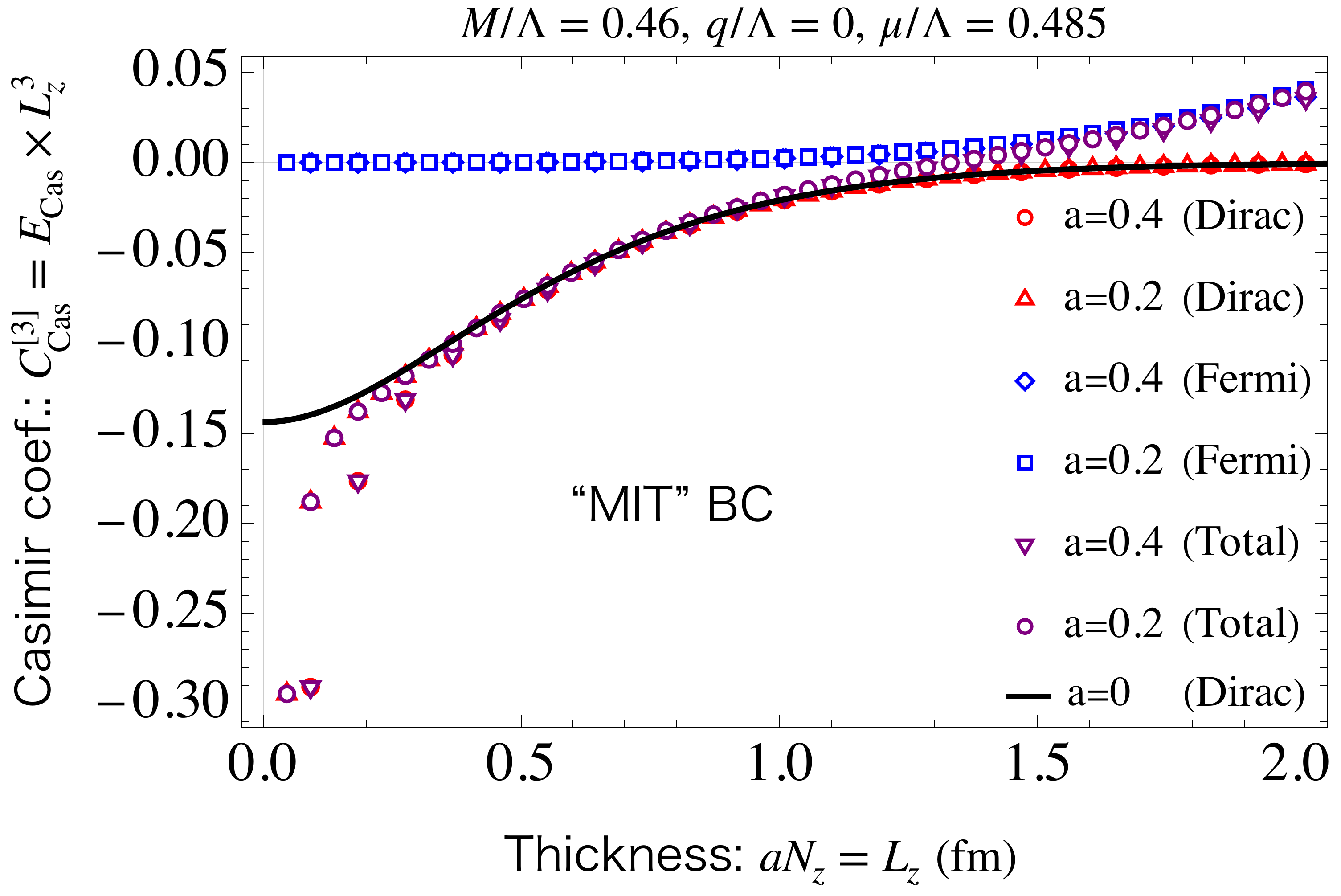}
\includegraphics[scale=0.15]{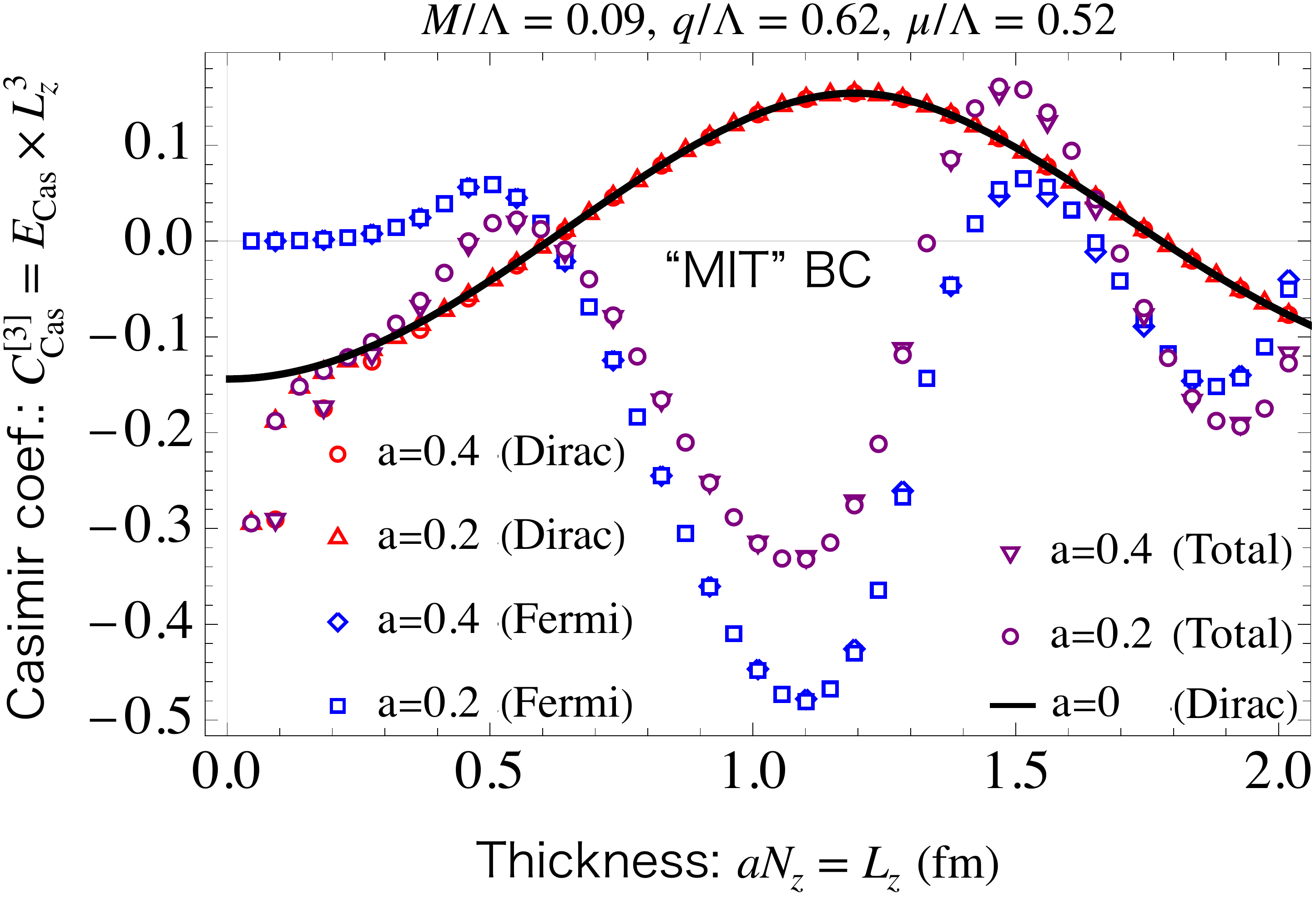}
\caption{\label{MITCasimir} Casimir coefficients in the zero-, intermediate-, and high-density regions with the ``MIT bag" boundary conditions.
We denote $a\Lambda$ as $a$ in the legends.}
\end{figure}

In this case, the infinite sum of eigenvalues is
\begin{align}
    \sum_{s=\pm}{\sum^\infty_{n=0}} \frac{\omega_{s,n}^{\rm ``MIT"}}{2},
\end{align}
and, the discretized $k_z$ is given as the zero point of the function 
\begin{align}
    \Delta^{\rm ``MIT"}_\pm(\omega)=1+e^{-2i\tilde{k}^{[\pm]}_zL_z}.
\end{align}
Using the derivation in Appendix~\ref{App:Lifshitz}, we obtain the Lifshitz formula with the ``MIT bag" boundaries: 
\begin{align}
E_{\rm Cas}^{\rm ``MIT"}
&=-2N_f N_c\sum_{s=\pm}\int_0 ^\infty\frac{d\xi}{2\pi}\int\frac{d^2k_\perp} {(2\pi)^2}\ln\big[1+e^{-2L_z\tilde{k}_z^{[s]}}\big], \notag\\
\tilde{k}^{[\pm]}_z&=\sqrt{k_\perp^2+M^2+\xi^2-\frac{q^2}{4}\mp iq\sqrt{k_\perp^2+\xi^2}}, \label{eq:ECas_MIT}
\end{align}
where $k_\perp^2 \equiv k_x^2+k_y^2$, and $d^2k_\perp \equiv dk_xdk_y$.
The overall minus sign and the factor of $+e^{-2L_z\tilde{k}_z^{[s]}}$ is a property of the ``MIT bag" boundary on fermion fields.
In the lattice regularization, we have to replace as $\sum^{\rm BZ}_{n} \to \frac{1}{2}\sum^{\rm BZ}_{n}$ in Eq.~(\ref{eq:E0sum}), and the sum is over $n=0,1,...,2N_z-1$.

Figure~\ref{MITCasimir} shows the results of the Casimir coefficients at zero-, intermediate-, and high-density regions.
In the case of the ``MIT bag" boundary, the number of discretized levels is twice larger than that with the PBC.
As a result, the period of oscillation is half of that with the PBC:
analogous to Eq.~(\ref{eq:period}),
\begin{align}
L_{z}^{\rm osc``MIT"} = \frac{\pi}{k_{\rm WP}}.\label{eq:period_MIT}
\end{align}

Furthermore, from these results, we can also discuss the relationship between the cases with the ``MIT bag" boundary and the APBC.
The discrete momenta are $k_z\to (n+1/2)\pi/L_z$ and $k_z\to (2n+1)\pi/L_z$, respectively.
For this reason, Eq.~(\ref{eq:ECas_MIT}) is equal to Eq.~(\ref{eq:ECas_APBC}) multiplying by $1/2$ and replacing as $L_z \to 2L_z$.
As a result, the Casimir energy with ``MIT bag" boundaries is exactly $16$ times smaller than that with the APBCs (the factor of $2^3$ is due to $3+1$ dimensions).
Therefore, using the results in Fig.~\ref{MITCasimir}, by multiplying the vertical axis by $16$ and the horizontal axis by $2$, we can obtain the exact results for the APBCs.

\bibliography{ref}

\end{document}